# Low latency global carbon budget reveals a continuous decline of the land carbon sink during the 2023/24 El Niño event

Piyu Ke[1,2], Philippe Ciais[3,*], Yitong Yao[4], Stephen Sitch[2], Wei Li[1], Yidi Xu[3], Xiaomeng Du[1], Xiaofan Gui[5], Ana Bastos[6], Sönke Zaehle[7], Ben Poulter[8], Thomas Colligan[8,9], Auke M. van der Woude[10], Wouter Peters[10], Zhu Liu[1], Zhe Jin[11], Xiangjun Tian[12], Yilong Wang[12], Junjie Liu[13], Sudhanshu Pandey[13], Chris O'Dell[14], Jiang Bian[5], Chuanlong Zhou[3], John Miller[15], Xin Lan[15,16], Michael O'Sullivan[2], Pierre Friedlingstein[2,17], Guido R. van der Werf[10], Glen P. Peters[18], Shilong Piao[11], Frédéric Chevallier[3]

1. Department of Earth System Science, Tsinghua University, Beijing 100084, China
2. Faculty of Environment, Science and Economy, University of Exeter, Exeter EX4 4QF, United Kingdom
3. Laboratoire des Sciences du Climat et de l'Environnement, University Paris Saclay CEA CNRS, Gif sur Yvette 91191, France
4. Department of Earth and Environmental Engineering, Columbia University, New York 10027, USA
5. Machine learning group, Microsoft research, Beijing 100080, China
6. Institute for Earth System Science and Remote Sensing, Leipzig University, Leipzig 04103, Germany
7. Department of Biogeochemical Integration, Max Planck Institute for Biogeochemistry, Jena 07745, Germany
8. Earth Sciences Division, NASA Goddard Space Flight Center, Greenbelt, MD 20771, USA
9. Earth System Science Interdisciplinary Center, University of Maryland, College Park, MD 20740, USA
10. Environmental Sciences Group, Dept of Meteorology and Air Quality, Wageningen University, Wageningen 6708 PB, the Netherlands
11. Institute of Carbon Neutrality, Sino-French Institute for Earth System Science, College of Urban and Environmental Sciences, Peking University, Beijing 100871, China
12. State Key Laboratory of Tibetan Plateau Earth System, Environment and Resources (TPESER), Institute of Tibetan Plateau Research, Chinese Academy of Sciences, Beijing 100101, China
13. Jet Propulsion Laboratory, California Institute of Technology, Pasadena 91011, CA, USA
14. Cooperative Institute for Research in the Atmosphere, Colorado State University, Fort Collins, CO 80523, USA
15. National Oceanic and Atmospheric Administration Global Monitoring Laboratory, CO 80303, USA
16. Cooperative Institute for Research in Environmental Sciences, University of Colorado Boulder, CO 80303, USA
17. Laboratoire de Météorologie Dynamique, IPSL, CNRS, ENS, Université PSL, Sorbonne Université, École Polytechnique, Paris 75005, France
18. CICERO Center for International Climate Research, Oslo 0349, Norway

***Corresponding author.** E-mail: philippe.ciais@cea.fr

# Abstract

The high growth rate of atmospheric $CO_2$ in 2023 was found to be caused by a severe reduction of the global net land carbon sink. Here we update the global $CO_2$ budget from January $1^{st}$ to July $1^{st}$ 2024, during which El Niño drought conditions continued to prevail in the Tropics but ceased by March 2024. We used three dynamic global vegetation models (DGVMs), machine learning emulators of ocean models, three atmospheric inversions driven by observations from the second Orbiting Carbon Observatory (OCO-2) satellite, and near-real-time fossil $CO_2$ emissions estimates. In a one-year period from July 2023 to July 2024 covering the El Niño 2023/24 event, we found a record-high $CO_2$ growth rate of $3.66 \pm 0.09$ ppm $yr^{-1}$ (± 1 standard deviation) since 1979. Yet, the $CO_2$ growth rate anomaly obtained after removing the long term trend is 1.1 ppm $yr^{-1}$, which is marginally smaller than the July-July growth rate anomalies of the two major previous El Niño events in 1997/98 and 2015/16. The atmospheric $CO_2$ growth rate anomaly was primarily driven by a 2.24 GtC $yr^{-1}$ reduction in the net land sink including 0.3 GtC $yr^{-1}$ of fire emissions, partly offset by a 0.38 GtC $yr^{-1}$ increase in the ocean sink relative to the 2015–2022 July-July mean. The tropics accounted for 97.5% of the land $CO_2$ flux anomaly, led by the Amazon (50.6%), central Africa (34%), and Southeast Asia (8.2%), with extra-tropical sources in South Africa and southern Brazil during April–July 2024. Our three DGVMs suggest greater tropical $CO_2$ losses in 2023/2024 than during the two previous large El Niño in 1997/98 and 2015/16, whereas inversions indicate losses more comparable to 2015/16. Overall, this update of the low latency budget highlights the impact of recent El Niño droughts in explaining the high $CO_2$ growth rate until July 2024.

# INTRODUCTION

The global $CO_2$ atmospheric growth rate in the decade 2013-2022 averaged $2.43 \pm 0.01$ ppm $yr^{-1}$. We previously reported a $2.82 \pm 0.08$ ppm $yr^{-1}$ growth rate in 2023 based on the global average of marine boundary layer sites (MBL) from the National Oceanic and Atmospheric Administration (NOAA) network, which has been updated by NOAA to $2.74 \pm 0.08$ ppm $yr^{-1}$ (https://gml.noaa.gov/ccgg/trends/gl_gr.html) [1–3]. The development of an El Niño in the second half of the year 2023 explained part of this high growth rate anomaly, notably with an extreme drought in the Amazon [3]. In this study, we extend this analysis to the first six months of 2024, when the El Niño continued until March [4]. The monthly annual $CO_2$ growth rate calculated as the difference between the monthly-smoothed $CO_2$ mole fraction at the MBL stations minus the one of twelve months before following the NOAA method, reached a peak of $3.67 \pm 0.13$ ppm $yr^{-1}$ in June 2024 (Supplementary Fig. 1a). The annual $CO_2$ growth rate calculated from June-July 2023 to June-July 2024 averaged $3.66 \pm 0.09$ ppm $yr^{-1}$ at the MBL stations compared to $2.46 \pm 0.01$ ppm $yr^{-1}$ from June-July to June-July averages during the decade 2013-2022 (Fig. 1a) [1,2]. The growth rate derived from independent OCO-2 satellite observations for June–July 2023 to June–July 2024 was $3.63 \pm 0.1$ ppm $yr^{-1}$ using the Growth Rates Using Satellite Observations data-driven approach (GRESO) from ref. [5]. The growth rate derived from our flux inversions of OCO-2 observations from July 2023 to July 2024 was $3.65 \pm 0.12$ ppm $yr^{-1}$, thus almost equal to GRESO (Supplementary section 1). The MBL and OCO-2 derived monthly growth rate time series shows peak values in September-November 2023 and July 2024, respectively (see Supplementary Figs. 1a and 2).

Differences between MBL and the Mauna Loa station (MLO) growth rates used in Ref. [3] are analyzed in the discussion. Overall, the MBL and the two OCO-2 derived growth rates show excellent agreement on annual scales, as shown in Fig 1a and Supplementary Fig. 1.

To update the global $CO_2$ budget until July $1^{st}$ 2024, we developed an integrated approach using top-down and bottom-up estimates of surface $CO_2$ exchanges. The annual $CO_2$ budget is presented here from July $1^{st}$ 2023 to July $1^{st}$ 2024, and fluxes are compared to previous years defined in the same way (Fig. 1b).

To update global fossil fuel and cement $CO_2$ emissions during the first half of 2024 (H1-2024) from July $1^{st}$ 2023 and July $1^{st}$ 2024, we used daily data from the Carbon Monitor project based on the collection of low latency activity data for six sectors [6–8] and averaged emissions estimates from the Global Carbon Project based on preliminary energy data with partial global coverage [9]. To allocate the annual emissions to the H1-2024 period, we used monthly emissions shares of annual emissions derived from the IEA-EDGAR monthly $CO_2$ emissions dataset [10]. The same monthly distribution as reported in 2023 was assumed for 2024 to estimate emissions for H1-2024. The relative uncertainties of fossil fuel and cement $CO_2$ emissions are within 10% for each dataset.

For the bottom-up global land carbon flux, we assessed the net land carbon flux which corresponds to the sum of anthropogenic land-use change emissions ($E_{LUC}$) and carbon uptake on other lands not affected by land-use change ($S_{LAND}$) in the Global Carbon Budget (see Methods). We used three dynamic global vegetation models (DGVMs) [11–15], OCN and ORCHIDEE-MICT previously used by Ref. [3], and LPJ-EOSIM [15]. The JULES model [16–18] used in Ref. [3] was excluded from this analysis based on the audit of the 2023 budget as this model gave a too high anomalous $CO_2$ loss, which was considered unrealistic. Net land $CO_2$ fluxes from the three DGVMs were driven by ERA5 atmospheric reanalysis forcing data until July $1^{st}$ 2024 at 0.5x0.5° resolution with the same protocol as in Ref. [3] (see Methods). Although we use only three DGVMs, their flux anomalies for previous years were close to the mean of all the models used in previous carbon budgets assessments [3,9,19], which gives us confidence that our small sample of low latency DGVMs can describe the net land carbon flux anomaly for the period from July 2023 to July 2024.

Our DGVMs simulate fire emissions as part of the net land $CO_2$ flux, in the range of 1.96 to 6.33 GtC yr$^{-1}$ from July 2023 to July 2024. This min-max range of DGVM-based fire emissions is much larger than that from observation-based products from the Global Fire Emissions Database (GFED4.1s) [20] and the Global Fire Assimilation System (GFAS, https://atmosphere.copernicus.eu/global-fire-monitoring) which are based on burned area and combustion energy observed by satellites, respectively. Our DGVMs simulate fires with prognostic fire modules using population density, lightning ignitions, climate and simulated fuel moisture but have weaknesses in capturing extreme forest fires and tropical forest degradation fires. Therefore, we corrected the DGVMs results by subtracting their original fire emissions and adding the mean of GFED4.1s and GFAS (see Methods). The GFED4.1s and GFAS fire emissions from July 2023 to July 2024 are 2.15 GtC yr$^{-1}$ and 1.88 GtC yr$^{-1}$ compared to 4.31 ± 2.2 GtC yr$^{-1}$ in the DGVMs during the same period.

In addition to the three low latency DGVMs listed above, we estimated the net land $CO_2$ flux using the Carbon Tracker Europe High-Resolution CTE-HR system [21] (0.1°×0.1° globally) with a spatially upscaled version of the SiB4 land $CO_2$ flux model forced with ERA5 atmospheric reanalysis [22]. The long-term mean Total Ecosystem Respiration is scaled so that the net ecosystem exchange (NEE) is equal to the long-term mean NEE of the Carbon Tracker Europe inversion, while GPP and respiration vary from hour–to–hour and across 10 plant-functional types. We also estimated the net land carbon flux using machine learning (ML) emulators of 19 DGVMs contributing to the Global Carbon Budget 2024 [9]. Each emulator was trained using as input features monthly gridded climate data, net $CO_2$ fluxes and annual vegetation maps of each native model until Dec. 2023 (Supplementary section 3). These emulators successfully reproduce the monthly $CO_2$ fluxes and the flux anomalies of the native DGVMs at 0.5 x 0.5° globally for test years that were not included in the training set (Supplementary Fig. 3). The net land carbon fluxes from CTE-HR and the 19 DGVM emulators were used here for the first time to predict $CO_2$ fluxes during H1-2024. Their results are compared to our three DGVMs as complementary information in this study (Supplementary Figs. 4 and 7).

For the bottom-up ocean carbon sink, we used machine learning emulators of each ocean biogeochemical and data-driven model as in Ref. [3] used for previous years in the global carbon budget assessments [23]. These ocean $CO_2$ flux emulators are trained by temporal trends and patterns of 16 original ocean models used by Global Carbon Budget 2022 [23] with input data being atmospheric $CO_2$ mixing ratio, sea surface temperature, ice cover, sea surface height, sea level pressure, sea surface salinity, mixed layer depth, wind speed and chlorophyll (see Methods and Supplementary section 2). After being trained until Dec. 2021, the emulators were run forward to make a projection of the monthly ocean sink at 1° spatial resolution for H1-2024, since only one ocean data-driven model provides low latency fluxes covering this period [24].

For the top-down budget, we used inversion models assimilating column-averaged $CO_2$ dry air mole fractions retrievals from the OCO-2 satellite (Level 4 ACOS V11 [25]) which cover H1-2024 while only a limited set of in-situ surface measurements is yet available for that period. Satellite observations from OCO-2 have been shown to provide similar skill in surface $CO_2$ flux estimation by inversions as assimilating the more accurate but sparse surface station measurements [26]. Moreover, the OCO-2 data have better coverage than the surface network across the tropics, which is an important advantage for separating $CO_2$ fluxes between the northern hemisphere and the tropics, and investigating flux anomalies between tropical continents, in particular between Amazon and South-east Asia affected by drought from July 2023 to July 2024. We used here three inversions: CAMS, CMS-Flux and GONGGA [26–29] assimilating the same OCO-2 column $CO_2$ mixing ratio data product [30,31]. The native spatial resolution of the CAMS monthly $CO_2$ fluxes is 90 km, which better matches that of our DGVMs (0.5°) and allows us to gain more insights into regional details of $CO_2$ fluxes without the usual smoothing effect of inversions. The spatial resolution of GONGGA is 2° by 2.5° (latitude×longitude) and the one of CMS-Flux is 4° by 5°. The inversion $CO_2$ fluxes were corrected for background natural fluxes related to the river loop of the carbon cycle as in Ref. [32], to provide anthropogenic carbon fluxes comparable to those simulated by bottom-up models.

The higher $CO_2$ growth rate from July 2023 to July 2024 echoes the impact of extreme warming, with each month warmer than the corresponding month in any previous year. The global temperature in the year 2024 was 0.72°C above the 1991–2020 average and 1.6°C warmer than the 1850–1900 pre-industrial level [33], with extreme heat waves in the northern mid latitudes and tropics, and drought in Central and South America, Central and southern Africa [34,35]. The period of H1-2024 marked the continuation of the El Niño that started in June 2023 and continued until March 2024 according to the Multivariate ENSO Index (MEI) from NOAA Physical Sciences Laboratory, with MEI > 0.5 indicating El Niño conditions (https://www.psl.noaa.gov/enso/mei) [4]. Overall the mean MEI of the 2023/24 El Niño was 0.71, that is a moderate El Niño compared to 2015/16 (MEI = 1.73) and to 1997/98 (MEI = 2.16). In the second half of 2023 (H2-2023) the Amazon experienced an extreme drought while Africa was on average wetter than usual [3].

## RESULTS AND DISCUSSION

### The global carbon budget from July 1$^{st}$ 2023 to June 30$^{th}$ 2024

We assessed the global yearly $CO_2$ budget for the period July 1$^{st}$ 2023 to June 30$^{th}$ 2024 (hereafter July-to-July) as it is not meaningful to provide a budget only during the first six months of the year 2024 when the seasonal uptake by the land is only half finished. The budget is usually based on calendar years, but here we assess the July-to-July period to reduce the latency and span the typical El Nino anomalies that are the main driver of year-to-year changes in global $CO_2$ fluxes.

Fig. 1a shows the June-July-to-June-July $CO_2$ growth rates (hereafter July-July) from MBL atmospheric stations. The $CO_2$ growth rate of $3.66 \pm 0.09$ ppm yr$^{-1}$ (mean ± 1 standard deviation) is a record-high value since 1979. Yet, the $CO_2$ growth rate anomaly obtained after removing the long term trend is of 1.1 ppm yr$^{-1}$, marginally lower than the July-July anomalies of the two previous El Niño events in 1997/98 and 2015/16. Comparing July-July growth rates between MBL stations and the longest single record of Mauna Loa shows differences ranging between 0.02 and 0.93 ppm yr$^{-1}$ among years since 1979, with a positive difference between MBL and Mauna Loa of 0.37 ppm yr$^{-1}$ between July 2023 and July 2024 (Supplementary Fig. 1b). Nevertheless, the cumulative increase of atmospheric $CO_2$ from July 1979 to July 2024 is similar within 1.85 ppm between Mauna Loa and MBL, indicating that the interannual differences between the two records balance over extended time periods.

Fig. 1b shows the bottom-up carbon budget, obtained from the mean of our two estimates of fossil emissions [6–9] combined with bottom-up net land $CO_2$ flux and ocean carbon fluxes. Fig 1c shows the top-down carbon budget based on the mean of the three OCO-2 inversions. Global fossil fuel and cement $CO_2$ emissions were 10.14 GtC yr$^{-1}$ (with a range of 10.14-10.15 GtC yr$^{-1}$ from our two estimates). In the bottom-up budget, the DGVMs estimated a net land source of $0.29 \pm 0.05$ GtC yr$^{-1}$ and in the top-down budget, the inversions gave a net land source of $0.09 \pm 0.26$ GtC yr$^{-1}$. The mean net land $CO_2$ flux of the DGVMs and inversions was a source of $0.19 \pm 0.13$ GtC yr$^{-1}$. The ocean $CO_2$ uptake from July 2023 to July 2024 was of $2.56 \pm 0.33$ GtC yr$^{-1}$ (ocean emulators: $2.87 \pm 0.56$ GtC yr$^{-1}$, inversions: $2.25 \pm 0.33$ GtC yr$^{-1}$) similar to the mean of the year 2023. The bottom-up budget imbalance, defined by the difference between fossil fuel emissions minus sinks from bottom-up models minus the observed $CO_2$ growth rate from July 2023 to July 2024, is of -0.2 GtC yr$^{-1}$ using the growth rate of MBL stations and a conversion factor of 2.124 GtC per ppm. Such a negative imbalance indicates that the bottom up models overestimate the global $CO_2$ sink during that period. The inversions on the other hand fit the $CO_2$ growth rate well and their imbalance is less than -0.05 GtC yr$^{-1}$ in July 2023 to July 2024 (Supplementary Fig. 1c). The imbalance of inversions during previous years is due to the fact that inversions prescribed slightly different emissions than those used in this study which are from the global carbon budget (GCB2024)

### Carbon fluxes from January 1$^{st}$ to June 30$^{th}$ 2024

From January 1st to June 30th 2024 (H1-2024) which covers the second year of the 2023/24 El Niño, global fossil emissions reached 5.12 GtC, an increase of 0.04 GtC (0.8%) compared to the same period of 2023. Although this increase is relatively modest, atmospheric $CO_2$ concentrations continued to rise at a fast rate, indicating a significant reduction in carbon sinks during this period.

The net land $CO_2$ flux during H1-2024 was a net uptake of 1.42 ± 2.06 GtC in the three DGVMs (note the large uncertainty) compared to 0.71 ± 0.42 GtC in the three inversions. We have a significant uptake as H1-2024 covers roughly half of the growing season in the northern hemisphere causing a strong periodical land uptake of $CO_2$, evidenced by the drawdown of the $CO_2$ atmospheric mixing ratio from April to mid-July at northern stations [36]. The Northern land regions, north of 30°N were a net carbon sink of 1.83 ± 1.94 GtC in the DGVMs and 1.18 ± 0.4 GtC in the inversions. On the other hand, the Tropical lands were a net source of 0.28 ± 0.23 GtC in the DGVMs and 0.49 ± 0.04 GtC in the inversions during H1-2024. Fire emissions were 0.68 GtC and 0.61 GtC during H1-2024, compared to 0.86 GtC and 0.7 GtC during H1-2023, respectively from GFED and GFAS.

The net land $CO_2$ flux anomaly during H1-2024 defined as the difference from previous H1 periods between 2015 and 2022 used as a reference, was a source of 0.74 ± 0.24 GtC in the three DGVMs. In comparison, the net land $CO_2$ flux anomaly was a small uptake of 0.13 ± 0.12 GtC during H1-2023 by the end of a wetter La Niña period, and a source of 1.5 ± 0.15 GtC during H2-2023 at the start of the El Niño period [3]. The net land $CO_2$ source anomaly of the DGVMs during H1-2024 represents a record high anomalous source compared to previous H1-periods since 2013 in the simulations from our three DGVMs. The land $CO_2$ flux anomaly was 0.29 ± 0.05 GtC in the OCO-2 inversions, thus smaller than the DGVM based anomaly.

The ocean sink was of 1.29 ± 0.13 GtC during H1-2024, in close agreement with the bottom-up models (1.51 ± 0.13 GtC). The ocean sink anomaly during H1-2024 is a sink of 0.2 ± 0.07 GtC in our bottom-up approach. Compared to the mean of the year 2023, the ocean sink during H1-2024 was 0.11 ± 0.03 GtC higher, mainly due to the continuing El Niño which decreased $CO_2$ sources in the Equatorial Pacific [24].

## Regional flux anomalies during the first half of 2024

To gain insights on which regions caused the large global $CO_2$ source anomaly in H1-2024, we analyzed the spatial patterns of flux anomalies for Jan-Mar (JFM) and Apr-Jun (AMJ) in the DGVM models and the inversions. The results are displayed in Fig. 2, in Supplementary Fig. 4 with Sib4 and the DGVM emulators, and in Supplementary Figs. 5-6 for the JFM and AMJ mean fluxes over the RECCAP2 regions along with their distributions relative to the same quarters in previous years.

Over the ocean, the most notable increases in carbon uptake were observed in the Pacific Ocean, consistent between the ocean models emulators and inversions (Fig. 2). Particularly, increased ocean $CO_2$ uptake was most pronounced in the eastern equatorial Pacific, consistent with suppressed upwelling of carbon rich waters during the developing El Niño[30]. The carbon sinks in the Arctic Ocean, the Atlantic Ocean, the Indian Ocean and coastal oceans remained relatively unchanged. There is a divergence in the Southern Ocean, where ocean model emulators suggest an increase, but the OCO-2 inversion indicates a decrease (Supplementary Figs. 6c-e and Supplementary Table 2).

Over the land, the regional net land $CO_2$ flux anomalies in Fig. 2 show dipoles of anomalous sources and sinks of similar magnitude between the DGVMs and the inversions, although the CAMS inversion has larger magnitudes for regional flux anomalies, possibly coming from its higher spatial resolution (Supplementary Fig. 7).

Over the northern lands (north of 30°N), the net land $CO_2$ flux anomaly was a source of $0.1 \pm 0.1$ GtC $yr^{-1}$ in the DGVMs and an uptake of $0.26 \pm 0.23$ GtC $yr^{-1}$ in the inversions during JFM-2024. The small magnitude of these JFM anomalies can be expected by the fact that soils are frozen and there is a minimum of vegetation activity over most of the northern lands. The northern net land flux anomaly in JFM is within 1-sigma of the variation of previous JFM periods during 2015-2022. During AMJ-2024, the northern land net $CO_2$ flux anomaly was $1.39 \pm 0.18$ GtC $yr^{-1}$ in the DGVMs (within 1-sigma of anomalies during 2015-2022) and $0.05 \pm 0.37$ GtC $yr^{-1}$ in the inversions (within 1-sigma of anomalies during 2015-2022). Such a large difference between DGVMs and inversions reflects differences for the spring and early summer northern $CO_2$ uptake: unlike inversions, the DGVMs are not constrained by the observed northern $CO_2$ atmospheric mixing ratio drawdown in AMJ. The inversions indicate an anomalous $CO_2$ uptake in AMJ-2024 in Eastern Europe, Central Asia, eastern Siberia and an anomalous $CO_2$ source in European Russia and Western Siberia, North America and China (Fig. 2). The DGVMs flux anomalies are consistent with inversions over boreal North America, European Russia and eastern Siberia.

In the second half of 2023 (H2-2023), we previously found anomalous sources of $CO_2$ of $0.14 \pm 0.05$ GtC in the Amazon, $0.21 \pm 0.07$ GtC in central Africa, and $0.06 \pm 0.03$ GtC in tropical Asia, respectively, using as reference period the mean all H2 periods during the period 2015-2022 covered by OCO-2 data. Updates of those regional $CO_2$ fluxes anomalies are provided in this study for the period H1-2024. Over the Tropics, we found net land $CO_2$ source anomalies during both JFM and AMJ-2024, with a fair spatial consistency between the mean of the DGVMs and the mean of the inversions (Fig. 2). Over tropical South America, anomalous $CO_2$ sources persisted during JFM-2024, but had a weaker magnitude than in late 2023 (see Fig. 2 of Ref. [3]). The DGVMs and the inversions locate anomalous sources over eastern Brazil, and the DGVMs further show anomalous sources over Southern Brazil. The magnitude of the $CO_2$ source anomaly over tropical South America north of 30°S is of $0.83 \pm 0.02$ GtC $yr^{-1}$ in the DGVMs and $0.13 \pm 0.14$ GtC $yr^{-1}$ in inversions during JFM-2024, smaller than the source anomalies during the preceding period of OND-2023 analyzed in Ref. [3]. This smaller net land $CO_2$ source anomaly is consistent with the fact that the drought in the Amazon peaked in late 2023 but diminished in JFM-2024 with only 7.5% of the region being under extreme drought (see Fig. 5). In AMJ-2024, the vanishing El Niño coincided with a return to net land $CO_2$ sink anomalies, but a source anomaly developed in Southern Brazil, with low rainfall. In tropical Africa, we found a large source anomaly in the DGVMs and a small source anomaly in the inversions in AMJ-2024, following a source anomaly of $0.21 \pm 0.37$ GtC $yr^{-1}$ in the DGVMs and $0.34 \pm 0.04$ GtC $yr^{-1}$ in inversions during JFM-2024. This transition to anomalous $CO_2$ sources in this region contrasts with the anomalous $CO_2$ sink that prevailed during DJF 2023 when conditions were wetter than normal [3]. In Southern Africa both in JFM and AMJ-2024 were net land $CO_2$ source anomalies while the Horn of Africa remained wetter and was a $CO_2$ sink anomaly. In tropical Asia and Oceania, both DGVMs and inversions diagnosed a net land $CO_2$ source anomaly in AMJ-2024 mainly over Southern China, continental Southeast Asia and Indonesia. In Northern Australia, despite continuing El Niño conditions usually associated with drought during the Australian summer and autumn, both DGVMs and inversions located an anomalous $CO_2$ uptake during AMJ-2024.

The monthly evolution of net land $CO_2$ fluxes anomalies in the tropics is displayed in Fig. 3a for the DGVMs and in Fig 3b for the inversions. The results indicate that the DGVMs show a sharp transition to a $CO_2$ source anomaly in August 2023 with a maximum loss anomaly in November 2023, mainly from the Amazon, a return to weaker anomalies in JFM-2024, and an emerging $CO_2$ source anomaly thereafter. ORCHIDEE, LPJ-EOSIM and OCN have a good agreement with each other (results with JULES in Supplementary Fig. 8). The inversions show earlier but smaller monthly net land $CO_2$ source anomalies than the DGVMs, starting in June 2023 and persisting until July 2024. The cumulative $CO_2$ source anomaly from June $1^{st}$ 2023 to July $1^{st}$ 2024 is $1.55 \pm 0.13$ GtC yr$^{-1}$ in the inversions compared to $2.25 \pm 0.35$ GtC yr$^{-1}$ in the DGVMs, suggesting that the DGVMs overestimated $CO_2$ losses during that period, possibly because they overestimate the response of photosynthesis to drought or because they ignore the transient storage in deadwood pools after drought induced mortality [37], as discussed in the section on model limitations below.

## Regional flux anomalies in relation to land water storage

The bivariate anomalies of net land $CO_2$ fluxes and total land water storage from the GRACE satellites assumed to represent soil water availability for plants, during each half-year period since Jan $1^{st}$ 2023 are shown in Fig. 4. There is a shift between H1-2023 and H2-2023 towards dryer anomalies associated with less $CO_2$ uptake or increased $CO_2$ sources (magenta areas in Fig 4b and 4c). From H1-2023 to H2-2024, we found a large increase in the extent of those types of dryer / $CO_2$ anomalous source anomalies from 23.0% to 37.2%, and then a decrease to 28.8% of the global land area from H2-2023 to H1-2024. In the northern extratropics, dryer conditions associated with decreased $CO_2$ uptake prevailed in North America, Central Asia and Western Siberia. In the Tropics, we observe a reduction of dryer / decreased $CO_2$ uptake anomalies over the Amazon in H1-2024 during the termination of the El Niño. However, the decrease in $CO_2$ uptake in Southern Brazil, which has been affected by a long-term drought, persists. We also observe a shift towards dryer / decreased $CO_2$ uptake in central Africa during H1-2024, as well as a persistence of drought in the Southern part of Africa. Eastern Africa remained under a regime of wetter / more $CO_2$ uptake between H2-2023 and H1-2024. In Asia and Oceania, the data in Fig 4a and 4b indicate $CO_2$ losses from inversions, yet associated with wetter conditions from GRACE. Northern Australia remained characterized by wetter / higher $CO_2$ uptake conditions.

## Comparison of the 2023/24 El Niño with the two previous episodes

We further analyzed the development of monthly water deficit anomalies and their impact on net land $CO_2$ fluxes for each tropical continent during the first and second years of the last three El Niño events in 1997/98, 2015/16 and 2023/24. The results displayed in Fig. 5 show that in the Amazon, water deficits during El Niño years usually develop during the dry season of the first El Nîño year, peak around January, and decrease in the second El Niño year. From MEI index values https://www.psl.noaa.gov/enso/mei, El Niño conditions typically terminate by April or May of the second year. In Southeast Asia, El Niño drought is synchronous with the Amazon. In central Africa, there is not necessarily a strong immediate teleconnection between El Niño and rainfall deficits, although El Niño years tend to be followed by dryer conditions. The delayed or weak response of African rainfall during El Niño years is due to interactions with Indian Ocean warming, seasonal effects, and atmospheric wave propagation [38].

During the 1997/98 El Niño, water deficits started in June 1997 and ended in April 1998, with a peak in September 1997 corresponding to a maximum of 43% of the Amazon region being under drought and 12% under extreme drought (see Methods for drought severity definitions based on water deficits). In Central Africa, there are two periodical dry seasons in Dec-Feb and Jun-Oct respectively (Fig. 5). In 1997/98, central African forests were only moderately affected by drought, with only 7.6% of their area under mild drought at peak and no severe drought conditions. The 2015/2016 El Niño was more severe and the area under drought in the Amazon was higher than in 1997/98 with 24% of the area being under extreme drought in Jan 2016, although the total duration of the water deficit period was shorter than in 1997/98 by 3 months. In Central Africa, the two dry seasons had clearly more severe droughts than in 1997/98, with up to 21.6% of the area being under extreme drought in the dry season of DJF 2016. In 2023/24, the El Niño event of interest in this study that has never been analyzed before, we can see in Fig 5 that the drought was much more severe than the two previous El Niño episodes. In the Amazon, there was a peak of areas under water deficit of 68.2% during the record drought of Sep-Dec 2023 [39–42]. The 2023/24 El Niño drought was severe during the second El NIño year in 2024 compared to previous second years in 1998 and 2016, and drought conditions restarted in June 2024 after the end of the El Niño (Fig 5).

Fig. 5 also shows anomalies of the net land $CO_2$ fluxes from the mean of the DGVMs during the last three El Niños. In general, the models simulated negative flux anomalies (anomalous losses, or reduced sinks) both during the onset of El Niño in the first year and its decay in the second year. Anomalous $CO_2$ losses are systematically more severe in the first year than in the second year over the Amazon, with a peak source anomaly during October when the usual dry season conjugates with the developing water deficit from El Niño. Generally, anomalous $CO_2$ losses are terminated in the Amazon by April of the second El Niño year. During the full El Niño periods of 1997/98, the DGVMs give a anomaly loss of $CO_2$ of $0.56 \pm 0.42$ GtC yr$^{-1}$ across the pan-tropics, an anomalous loss of $0.56 \pm 0.2$ GtC yr$^{-1}$ in the Amazon, an anomalous uptake of $0.26 \pm 0.06$ GtC yr$^{-1}$ in Central Africa and a anomalous loss of $0.28 \pm 0.14$ GtC yr$^{-1}$ in South-east Asia (Fig. 5). In 2015/16, the anomalous pan-tropical loss was larger than in 1997/98, with a value of $1.58 \pm 0.03$ GtC yr$^{-1}$, including anomalous losses of $0.55 \pm 0.06$ GtC yr$^{-1}$ in the Amazon, $0.41 \pm 0.03$ GtC yr$^{-1}$ in Central Africa and $0.29 \pm 0.06$ GtC yr$^{-1}$ in South east Asia, respectively. In 2023/24, the pan tropical loss reached up to $2.31 \pm 0.14$ GtC yr$^{-1}$, with a loss of $1.51 \pm 0.13$ GtC yr$^{-1}$ in the Amazon, $0.55 \pm 0.12$ GtC yr$^{-1}$ in Central Africa and $0.11 \pm 0.03$ GtC yr$^{-1}$ in Southeast Asia, respectively. This result from DGVMs suggesting a more severe $CO_2$ loss anomaly during the 2023/24 event compared to the two previous events in 1997/98 and 2015/16 is not corroborated by inversions results in Fig. 3 c-d which tend to show a similar $CO_2$ loss anomaly between the 2023/24 event and the 2015/16 El Niños. Note that the OCO-2 inversions starting in 2015 do not cover the 1997/98 El Niño.

Fire emissions anomalies were particularly severe in the Amazon region, closely aligned with periods of intense drought described above. During H2-2023, extensive drought conditions in the Amazon led to significant wildfire emissions, with unprecedented carbon emissions particularly noted in Brazil and Venezuela. Fire emissions further increased into H1-2024, resulting in record-breaking wildfire emission anomalies (0.06 GtC relative to 2015-2022 average) across northern South America, notably Bolivia, Guyana, and Suriname, with Bolivia recording its highest fire emissions since 2003. In contrast, Southeast Asia and Central Africa exhibited relatively modest fire emission anomalies during the same period. Southeast Asia, in particular, saw fire emissions remaining stable or even below average levels in H1-2024, suggesting effective regional fire management or milder climatic conditions. Central Africa similarly displayed relatively stable fire emission patterns throughout the 2023/24 El Niño event, consistent with the less pronounced drought anomalies discussed previously [43].

## Agreement between models and models limitations

The spatial contrasts of the net land CO2 fluxes anomalies shown in Fig. 2 are not consistent at a resolution of 1°x1° in JFM (spatial $R^2 = 0.034$) but have a better agreement in AMJ ($R^2 = 0.2$). Here, the inversion fluxes for GONGGA and CMS-flux have been re-interpolated to 1°x1° from their coarser native resolution. We also analyzed the level of agreement between the different models used in this study (Supplementary Fig. 7). The level of agreement between the DGVMs in each 1° grid cell is defined as good if the three models have a JFM or AMJ flux anomaly higher than 1-sigma from the 2015-2022 period in this grid cell and its magnitude differs by less than 30%. We found that 80.24% and 80.71% of the northern lands and 68.84% and 88.94% of the tropics have a good agreement for the JFM and AMJ of 2024. At a larger scale of continents, the agreement of flux anomalies was also good between the models. The level of agreement between the three inversions was considered good if 2 models out of three have a flux anomaly that differs by less than 20%. At the scale of the coarsest inversion model, CMS-Flux (with a resolution of 4° × 5°), the agreement of inversions over the northern lands was 90.44% and 83.97% for JFM and AMJ of 2024, respectively. Similarly, over the tropical region, inversion agreement reached 86.54% in JFM and 83.64% in AMJ, indicating consistently high levels of agreement across different latitude bands.

A main limitation of the DGVM models used in this study for studying the response of $CO_2$ fluxes to drought is that they have no drought induced tree mortality parameterisation, a limitation common to most other DGVMs. DGVMs simulate a $CO_2$ loss from a deficit in gross primary productivity caused by water stress, and generally smaller changes in plant and soil $CO_2$ respiration since warmer conditions increase respiration rates, while dryer conditions and coincident drought-induced decreases of GPP reduce them [44]. Therefore, even though the net land $CO_2$ flux anomalies during 2023/24 simulated by our DGVMs are qualitatively consistent with the independent anomalies from inversions, DGVM models do not simulate lagged mortality processes as observed in the Amazon and in other tropical wet forest biomes [45–47]. Increased mortality also causes a transient increase of coarse woody debris on forest floor which will not be decomposed immediately and will cause a temporary additional carbon storage typically during 20 years [37,48] with a lagged release of $CO_2$ to the atmosphere, contributing to a higher atmospheric $CO_2$ growth rate in the following decades. Since DGVMs do not simulate increased tree mortality during drought followed by transient storage in deadwood, this important process controlling $CO_2$ fluxes is absent from our budget assessment, which may explain why the DGVMs $CO_2$ source anomaly in 2023/24 is larger than that of inversions, as noted by ref [37]. Last, our DGVMs were corrected for fire emissions using data driven emission products based on moderate resolution sensors that underestimate small degradation fires in wet tropical forests [49]. It is expected that new burned area products with more small fires [50] will increase emissions when emissions datasets will be available from those products.

The three inversions used different priors and transport models, and different aggregations of OCO-2 observations into atmospheric model grid boxes. The spread of the inversions is of 0.51 GtC yr$^{-1}$ for global $CO_2$ fluxes during H1-2024, indicating that the budget over only six months is relatively well constrained by OCO-2 observations. The spread of inversions was 0.28 GtC for the ocean sink and 0.43 GtC for the land sink during H1-2024, which is larger than the spread of their annual land sinks from July 2023 to July 2024, but smaller than the spread of their annual ocean sinks. This indicates that inversions have a higher uncertainty when land budgets are analyzed on a semester basis. The spread of our inversions is also lower than that of the three DGVM models. We found that the three inversions agree about the sign of tropical flux anomalies in each continent, and on the spatial patterns of flux anomalies during H1-2024 (spatial $R^2$ = 0.05 for JFM and 0.01 for AMJ between the three models at 4°x5° resolution, the resolution of the coarsest inversion - CMS-Flux). However, the magnitude of regional anomalies in CAMS is twice larger than in CMS-Flux and GONGGA, which reflects its higher spatial resolution. This difference needs to be investigated in further studies.

# An audit of our latency models against multi-model ensembles from the Global Carbon Budget in 2023

In this study, to be consistent with GCB2024, we only use the growth rate from MBL stations. In our previous low latency budget [3], we used the growth rate from MBL stations as in the previous global carbon budget assessment (GCB2024) but also from the Mauna Loa (MLO) station. The growth rate at MLO was of 0.37 ppm yr$^{-1}$ higher than the MBL stations from June-July 2023 to June July 2024 according to NOAA Global Monitoring Laboratory data (https://gml.noaa.gov/ccgg/trends/mlo.html; see Supplementary Figure 1b). If the budget imbalance from July 2023 to July 2024 would be calculated using the MLO station, the imbalance would be 0.58 GtC yr$^{-1}$ instead of -0.2 GtC yr$^{-1}$ with MBL data, implying a missing source in our bottom-up budget instead of a small missing sink with MBL data (Supplementary Fig. 1c). The monthly MBL minus Mauna Loa growth rate difference took its largest negative value of 1.14 ppm yr$^{-1}$ in February 2024 (Supplementary Fig. 1a). This persistent and intriguing positive difference between MLO and MBL growth rates could be due to changes in emissions and regional atmospheric transport impacting MLO, or possibly due to volcanic activity at MLO between Dec. 2022 and July 2023 that resulted in moving the instrument to nearby Mauna Kea (https://gml.noaa.gov/ccgg/trends/). Cumulatively, over a long enough time, the difference of atmospheric $CO_2$ increase eventually cancels out between MLO and MBL, as shown above.

For this second edition of the low latency global and regional $CO_2$ budget, we benchmarked our previous results for the year 2023 based on the JULES, OCN, ORCHIDEE DGVMs and the CAMS inversion (published in October 2024) [3] against the mean of all the GCB2024 DGVMs and inversions in Ref. [9] (published in March 2025). Note that the net land $CO_2$ flux of each DGVM was simulated from climate forcing only but adjusted to the mean of the TRENDY models S3 simulation in the period 2019-2022 which includes land use change. Therefore, our models include implicitly the average flux from land use change but ignore recent variations of land use change in 2023 and 2024. The results of the comparison are shown in Supplementary Figures 9a-b for the DGVMs and inversion estimates of the net land $CO_2$ flux, and in Supplementary Figures 9c-d for ocean model emulators and inversions over the ocean. The CAMS inversion that was used in Ref. [3] is very close to the median of the 8 satellite and 6 in situ inversions used by GCB2024. On the other hand, the mean of the three DGVMs used by Ref. [3] (JULES, ORCHIDEE, OCN) shows a lower net land $CO_2$ flux than the mean of the 20 DGVMs from GCB2024, by 1.07 GtC yr$^{-1}$. Our lower DGVM net land flux is still in the range of GCB2024, with 6 models producing sinks of the same magnitude as the three DGVMs of Ref. [3]. Inspection of the results of each of our three DGVMs show clearly that JULES had a very high sensitivity to drought in the Tropics and simulates larger fire emissions durga dry periods (Supplementary Figure 9e) and therefore produced a lower net land $CO_2$ sink in 2023 than our two other models of Ref. [3], and the lowest sink of all the GCB2024 DGVMs (Supplementary Figure 7). This is the reason why the JULES results were excluded from this new assessment. Values of global, regional, anomalies of land $CO_2$ fluxes with and without JULES are shown in

Supplementary Table 1. For the ocean sink which has less variability from one year to another than the land $CO_2$ sink, the results of our 16 emulators are extremely close to those of ocean biogeochemistry models and data driven models used by GCB2024. Similarly, the results of the CAMS inversion for the ocean sink in 2023 is very close to the mean of all the inversions used by GCB2024.

## DGVM emulators to predict land $CO_2$ fluxes in the first half of 2024

In this study, for the first time, we developed 19 ML emulators of DGVMs used in GCB2024. These emulators are deep learning models trained to simulate the monthly net land sink from the simulation 2 of each original GCB2024 DGVM at 0.5x0.5° resolution using as input features climate, the vegetation map of each original model, and the previous 12 months of $CO_2$ fluxes of each model. The training / validation period of the emulators is from 2000 to 2023 and the prediction period is H1-2024. The emulators predicted fluxes and fluxes anomalies at 0.5° that are very close to the original models during years used as test unseen by the training (spatial $R^2$ in the range 0.08-0.35 at 0.5°x0.5° across different quarters and models; see flux anomalies maps for three GCB2024 models during test years in Supplementary Fig. 3). In their prediction for the H1-2024 period, the mean of the emulators gives a global sink of 2.05 ± 0.37 GtC and a source anomaly of 0.21 ± 0.33 GtC. The monthly anomalies of tropical net land $CO_2$ fluxes from the mean of the 19 emulators during H1-2024 is given in Fig. 6 (supplementary Fig. 8 with JULES included). The result shown in Fig. 6a indicates that the emulators of the three DGVMs models used in this study, ORCHIDEE, OCN and LPJ-wsl, give a similar temporal evolution of tropical $CO_2$ flux anomalies, but smaller anomalous $CO_2$ losses than the low latency native models. This difference is possibly due to the fact that the GCB2024 models used to train emulators were run with the CRU JRA atmospheric forcing whereas our three low latency DGVMs shown in Fig. 6 were run using the ERA5 forcing. Fig. 6b displays the monthly net land $CO_2$ flux anomalies from the 19 emulators (mean and standard deviation). The results show that the mean of the 19 emulators has a smaller loss anomaly than our three low latency DGVMs, and that the mean of the emulators is better aligned with inversions (Fig. 3b). The performances of the emulators will continue to be assessed at least for one year against the results of each native DGVM in the next edition of the global budget. Then, emulators will be used as part of the next edition of our low latency assessment of the $CO_2$ budget.

## Conclusions

Overall, we found a record high $CO_2$ growth rate from June-July 2023 to June-July 2024 associated with a slight increase of the ocean sink and to a strong reduction of the net land $CO_2$ flux to a value close to no net land $CO_2$ source (0.09 ± 0.26 GtC yr$^{-1}$) in the inversions which provide fluxes that match the observed $CO_2$ growth rate. Our three low latency DGVMs give a larger net land $CO_2$ source of 0.29 ± 0.05 GtC yr$^{-1}$. In the inversions, the net tropical land $CO_2$ source anomaly from July 2023 to July 2024 was 0.85 ± 0.08 GtC yr$^{-1}$, compared to 2.18 ± 0.09 GtC yr$^{-1}$ in DGVMs. In the inversions, the 2023/24 net land $CO_2$ source anomaly was similar in magnitude to the one diagnosed for the previous El Niño of 2015/16 while the 2023/24 source anomaly was more severe in the DGVMs. Our DGVM results indicate that the Amazon experienced a less severe drought-induced $CO_2$ loss in H1-2024 compared to the previous half year of H2-2023, while central Africa had a very severe drought after April 2024 during the second dry season in this region, causing new $CO_2$ losses. Continental and maritime Southeast Asia also showed a $CO_2$ source anomaly in the DGVMs despite a relatively moderate drought in H1-2024. We tested for the first time ML emulators of the 19 DGVMs used in the global carbon budget trained to reproduce the fluxes of the native models until December 2023 and used for predicting the land CO2 fluxes during H1-2024. The mean of the emulators give a net tropical land $CO_2$ source anomaly close to the one simulated by inversions, which gives us confidence to use emulators in subsequent assessments. It is interesting to see that the Amazon experienced a second extreme drought during the dry season in H2-2024 while central Africa continued to be under drought, an important event that will be the focus of the following low-latency budget assessment.

# METHODS

**Atmospheric $CO_2$ growth rate (CGR).** We used the monthly time series of globally averaged marine surface (MBL) atmospheric $CO_2$ dry air mole fraction covering the period from January 1979 to November 2024, and the MLO station data from March 1958 to February 2025, both provided by NOAA's Global Monitoring Laboratory (NOAA/GML)[1,2]. The July-July annual MBL growth rate is determined by averaging the most recent June and July months, corrected for the average seasonal cycle, and subtracting the average of the same period in the previous year (https://gml.noaa.gov/ccgg/trends/gl_gr.html). For the MLO data, the July-July annual mean growth rate is estimated using the average of the most recent May-August months, corrected for the average seasonal cycle, and subtracting the same four-month average of the previous year centered on July 1st (https://gml.noaa.gov/ccgg/trends/gr.html).

**Global fire $CO_2$ emissions.** Both the Global Fire Emissions Database version 4.1 including small fire burned area (GFED4.1s)[20] and the Global Fire Assimilation System (GFAS, https://atmosphere.copernicus.eu/global-fire-monitoring) from Copernicus Atmosphere Monitoring Service (CAMS) are used to derive monthly global fire $CO_2$ emissions. GFED4.1s combines satellite information on fire activity and vegetation productivity to estimate the gridded monthly burned area and fire emissions, and has a spatial resolution of 0.25° × 0.25°[20]. Note that GFED4.1s fire emissions for the 2017 and onwards period are from the beta version. The GFAS assimilates fire radiative power (FRP) observations from satellites to produce daily estimates of biomass burning emissions. We aggregated the GFAS daily fire emissions into monthly emissions.

**Terrestrial $CO_2$ fluxes.** Terrestrial carbon fluxes are derived from the mean of three Dynamic Global Vegetation Models (DGVMs), specifically ORCHIDEE, OCN and LPJ-EOSIM. The methodology for estimating terrestrial carbon fluxes for the period 2010-2023 aligns with the TRENDY protocol, used in Global Carbon Budget[51], albeit with modifications due to the use of ERA5 climate forcing, and the fact that land-use forcing is not updated with such short latency. ERA5 forcing has been available since 1940[22], but preliminary simulations with DGVMs showed some issues with precipitation forcing before 1960. The simulations performed here correspond to the S2 experiment, starting from steady-state in 1960, with time varying climate and $CO_2$ forcing, and land-use fixed at 2010, as described in Ref. [52].

Since the model simulations start from a steady-state in the 1960s, they cannot account for the carbon balance before the industrial revolution, which leads to an underestimation of the mean trend compared to standard DGVMs. Therefore, we calibrated them using the 2019-2022 TRENDY DGVMs simulation 3 from the Global Carbon Budget 2023 [19]. This was done by calculating for each grid cell the median of the carbon flux from the TRENDY models during 2019-2022 and the mean of each of the DGVMs used in this study, then subtracting from each DGVM on each grid the difference so that they match the median flux of the TRENDY models. In other words, our DGVMs are used for predicting interannual anomalies of $CO_2$ fluxes.

**Ocean $CO_2$ fluxes.** Ocean carbon fluxes are derived from a suite of emulators based on both biogeochemical models and data-driven models. We updated estimates from 8 Global Ocean Biogeochemical Models and 8 data products included in the Global Carbon Budget 2022 to create a near-real-time framework [53]. This update employs Convolutional Neural Networks (CNNs) and semi-supervised learning techniques to capture the non-linear relationships between model or product estimates and observed predictors. As a result, we obtain a near-real-time, monthly grid-based dataset of global surface ocean fugacity of $CO_2$ and ocean-atmosphere $CO_2$ flux data, extending to December 2023. More details are given in Ref. [53] and in Supplementary section 2.

**Anthropogenic $CO_2$ emissions.** For the period from 2010 to 2023, we used global fossil fuel and cement $CO_2$ emissions estimates from the latest edition of the global carbon budget[9]. For 2024, we used the average of emissions from the Carbon Monitor project based on near-real-time activity data from 6 sectors [6–8] and from an updated estimate using the same methodology that the global Carbon Budget [9], i.e. based on energy data available by the time of the publication and projections for countries with no available data. The Carbon Monitor data give a global emission of +0.88% compared to 2023 while the Global Carbon Budget approach gives a global emission of +0.8% compared to 2023. We then applied monthly emissions proportions derived from the IEA-EDGAR monthly $CO_2$ emissions dataset from Emissions Database for Global Atmospheric Research (EDGAR) [10] to partition annual emissions into the first and second halves of each year. The same monthly distribution observed in 2023 was assumed for 2024 to estimate emissions specifically for H1-2024.

**Atmospheric inversions.** We used three atmospheric inversion driven by the OCO-2 satellite atmospheric $CO_2$ column-average dry air mole fraction data, which all contributed to the annual $CO_2$ budget assessment. CAMS has a net land $CO_2$ flux over the period 2015-2022 being a sink of 1.93 GtC yr$^{-1}$, CMS-Flux has a smaller mean land $CO_2$ flux of 0.94 GtC yr$^{-1}$ and GONGGA of 1.24 GtC yr$^{-1}$. CAMS has a smaller mean net land $CO_2$ flux being a sink in the northern extratropics (> 30°N) of 1.51 GtC yr$^{-1}$, CMS-Flux of 1.88 GtC yr$^{-1}$, and GONGGA of 1.56 GtC yr$^{-1}$. All inversions have very similar year-on-year anomalies of land $CO_2$ fluxes since 2015.

**CAMS inversion** This system provides global estimates of weekly greenhouse gas fluxes with a typical 4-month latency, now at a resolution of about 90 km based on a hexagonal-pentagonal mesh of the globe. The product used here is version FT24r2. It followed the usual production and quality control process of the CAMS products. It covers the OCO-2 period from October 2014 to September 2024, and its mean fluxes and anomalies are close to the median of inversions used in previous assessments[9,19] (see Supplementary Fig. 9c, d). The underlying transport model was nudged towards horizontal winds from the ERA5 reanalysis. The inferred fluxes were estimated in each horizontal grid point of the transport model with a temporal resolution of 8 days, separately for day-time and night-time. The prior values of the fluxes combine estimates of (i) gridded monthly fossil fuel and cement emissions (GCP-GridFED version 2024.0[54]) extended to year 2024 following Chevallier et al. (2020)[27] using the emission changes reported by https://carbonmonitor.org/, together with anomalies in retrievals of $NO_2$ columns from the Tropospheric Monitoring Instrument (TROPOMI, offline and processing and RPRO when available, van Geffen et al., 2019[55]), (ii) monthly ocean fluxes (Chau et al. 2024a[24], 2024b[56]), 3-hourly (when available) or monthly biomass burning emissions (GFAS) and climatological 3-hourly biosphere-atmosphere fluxes taken as the 1981-2020 mean of a simulation of the ORganizing Carbon and Hydrology In Dynamic

EcosystEms model, version 2.2, revision 7262 (ORCHIDEE, Krinner et al. 2005[13]). The variational inversion accounts for spatial and temporal correlations of the prior errors, resulting in a total 1-sigma uncertainty for the prior fluxes over a full year of 3.0 GtC·yr$^{-1}$ for the land pixels and of 0.2 GtC·yr$^{-1}$ for the marine pixels.

**GONNGA inversion** This system provides global estimates of $CO_2$ fluxes over the period of 2015-2024 at a spatial resolution of 2° by 2.5° (latitude×longitude) with 47 layers in the vertical direction from the surface to the top of the atmosphere. It follows the production and quality control process as described in Ref. [28]. The model is driven by Modern-Era Retrospective analysis for Research and Applications 2 (MERRA-2) meteorological data. The 3-hour fluxes were optimized in each horizontal grid point of the transport model in each 14-day temporal window. The prior values of the fluxes combine estimates of (i) gridded monthly fossil fuel and cement emissions (GCP-GridFED version 2024.0[54]) extended to year 2024 using the Global gRidded dAily $CO_2$ Emission Dataset-GRACED (https://carbonmonitor-graced.com/) [57]; (ii) gridded monthly ocean flux from CarbonTracker 2022 $pCO_2$-Clim prior data, which were derived from the Takahashi et al. (2009) climatology of seawater $pCO_2$; (iii) gridded monthly biosphere-atmosphere $CO_2$ exchange fluxes simulated by ORCHIDEE-MICT model extended to year 2024 by repeating the fluxes in 2023; (iv) gridded monthly biomass burning emissions from the Global Fire Emissions Database (GFED; version 4.1s). The prior errors account for spatial and temporal correlations, resulting in a total 1-sigma uncertainty for the prior global fluxes over a full year of 4.7 GtC·yr$^{-1}$ for land fluxes and of 0.3 GtC·yr$^{-1}$ for ocean fluxes.

**CMS-Flux inversion** This system provides global estimates of $CO_2$ fluxes over the period of Jan 2010- July 2024 at a spatial resolution of 4° by 5° (latitude×longitude) with 47 layers in the vertical direction from the surface to the top of the atmosphere. The inversion setups follow ref. [29] . The model is driven by Modern-Era Retrospective analysis for Research and Applications 2 (MERRA-2) meteorological data. CMS-Flux optimizes ocean and land fluxes at monthly temporal resolution, assuming known diurnal cycle from and the prior land and ocean carbon fluxes. The prior fluxes include (i) gridded monthly fossil fuel and cement emissions (GCP-GridFED version 2024.0[54]) extended to year 2024 using the Global gRidded dAily $CO_2$ Emission Dataset-GRACED (https://carbonmonitor-graced.com/) [57]; (ii) 3-hourly land carbon fluxes including fire emissions from CARDAMOM; (iii) 3-hourly ocean fluxes from either ECCO-Darwin or daily ocean fluxes from MOM-6. We carried out two sets of top-down flux inversions with two different ocean prior fluxes, and the final results are the mean of the two flux inversions. CMS-Flux assumes no spatial and temporal correlations in the prior flux uncertainties. The 1-sigma uncertainties for land and ocean fluxes are 1.3 GtC·yr$^{-1}$ and 0.3 GtC·yr$^{-1}$ respectively.

**Drought characteristics**
The characteristics of meteorological drought in the tropics is described by the Cumulative Water Deficit (CWD). Based on or precipitation data from ERA5-Land, we calculated monthly CWD for each pixel, as :

$$CWD_m = \min(0, CWD_{m-1} + P_m - 100) \text{ if } P_m < 100, \text{ else } CWD_m = 0 \quad \text{Eq. (1)}$$

with $m$ being the month within the hydrological year spanning from October of the previous year to September of the current year. $CWD_m$ is the CWD of the current month, which is equal to the CWD from previous months ($CWD_{m-1}$) plus the difference between the precipitation of the current month ($P_m$) and ET (assumed to be 100 mm). The water deficit accumulates over the entire hydrological year.

If monthly rainfall falls below 100 mm, the forest is considered to be experiencing a water deficit. For each grid cell, the Z score of CWD at the i$^{th}$ grid for the m$^{th}$ month is calculated as:

$$Z_{CWD,m,i} = \frac{CWD_{m,i} - \mu_{CWD,m,i}}{\sigma_{CWD,m,i}} \quad \text{Eq. (2)}$$

where $\mu_{CWD,m,i}$ and $\sigma_{CWD,m,i}$ are the mean value and standard deviation of CWD at the i$^{th}$ grid for the m$^{th}$ month during the reference period. We used the monthly $Z_{CWD}$ to identify the spatial extent, duration, severity, onset, and end of the El Nino related drought events in the tropical forests. Drought intensity was classified into five categories, slight ($Z \in [-1,0)$), mild ($Z \in [-1.645, -1)$), moderate ($Z \in [-1.96, -1.645)$), severe ($Z \in [-2.576, -1.96)$), and extreme ($Z < -2.576$).

## Data availability

The data from Global Carbon Budget 2024 are available at www.icos-cp.eu/science-and-impact/global-carbon-budget/2024. The OCO-2 retrievals are available at disc.gsfc.nasa.gov/datasets?page=1&keywords=OCO-2. NOAA/GML CO2 data are available at https://gml.noaa.gov/ccgg/trends/global.html. The Carbon Monitor fossil emissions dataset is available at carbonmonitor.org. The GFED 4.1s fire emissions dataset is available at geo.vu.nl/~gwerf/GFED/GFED4/. The GFAS fire emissions dataset is available at atmosphere.copernicus.eu/global-fire-monitoring/. The ERA5 monthly averaged data is available at cds.climate.copernicus.eu/cdsapp#!/dataset/reanalysis-era5-single-levels-monthly-means?tab=overview. The Multivariate ENSO index is available at www.psl.noaa.gov/enso/mei. The GRACE/FO TWS data used in this study are available at www2.csr.utexas.edu/grace/RL06_mascons.html.

## Author contributions

P.C. and P.K. designed the research; P.K. and P.C. performed the analysis; P.K., P.C. and Y.Y. collected and analyzed the research data; P.C. and P.K. created the first draft of the paper; all authors contributed to the interpretation of the results and to the text.

**Conflict of interest statement.** None declared.

## Acknowledgements

The authors acknowledge support from the National Key R&D Program of China (2023YFE0113000), the Carbon Neutrality and Energy System Transformation (CNEST) Program led by Tsinghua University, and the International Joint Mission on Climate Change and Carbon Neutrality. P.C., F.C., A.B., S.S., P.F. acknowledge support from the European Space Agency Climate Space RECCAP2-CS project (ESA ESRIN/4000144908), ESA Carbon-RO (4000140982/23/I-EF) and the CALIPSO project funded by the generosity of Schmidt Science. P.C., G.P.P. received funding from the European Union's Horizon Europe research and innovation programme under grant agreement No 101081395 (EYE-CLIMA). The CAMS simulations were granted access to the HPC resources of TGCC under the allocation A0170102201 made by GENCI. A.B. thanks Evgenii Churiulin for support in setting up the OCN model simulations. Work of J.L. and S.P. was conducted at the Jet Propulsion Laboratory, California Institute of Technology, under a contract with the National Aeronautics and Space Administration (80NM0018D0004). Additionally, J.L. acknowledges the funding support from NASA Orbiting Carbon Observatory Science Team program. AvdW acknowledges Remco de Kok, Ingrid Luijkx and Joram Hooghiem for their contributions to CTE-HR and the ICOS-CP for hosting CTE-HR data. This work was also supported by a computing grant from the Dutch national e-infrastructure with the support of the SURF Cooperative (NWO-2023.003). The authors also acknowledge sponsorship from Microsoft Research Asia (MSRA). This research was supported in part by the NOAA cooperative agreement NA22OAR4320151. The statements, findings, conclusions, and recommendations are those of the author(s) and do not necessarily reflect the views of NOAA or the U.S. Department of Commerce.

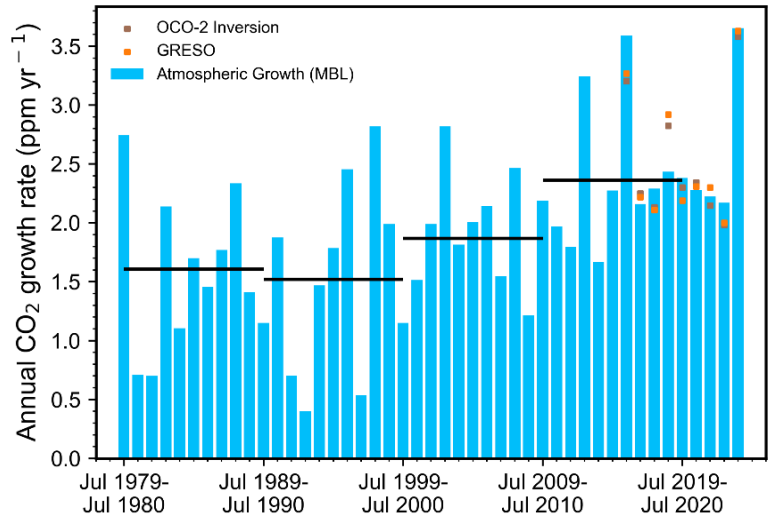
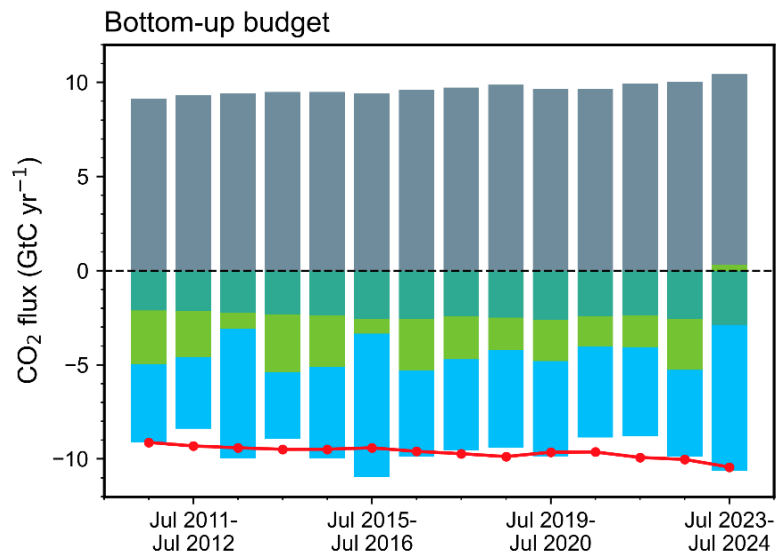
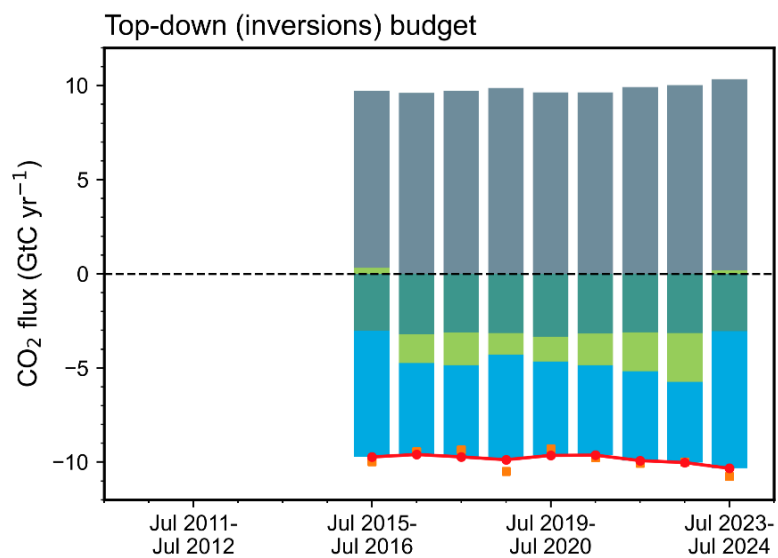
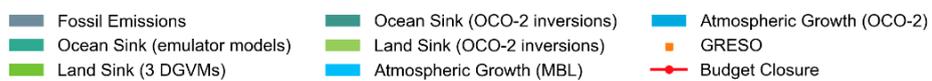

**Fig. 1 Atmospheric $CO_2$ growth rate (1979–2024) and carbon budget (2010–2024), calculated for July–July annual periods.** (a) Annual June-July to June-July growth rate from marine boundary layer surface stations (MBL, blue bars), OCO-2 (dark brown squares) and the Growth Rates Using Satellite Observations approach from ref. [5] (GRESO, orange squares). Our analysis is based on MBL. MBL is the average of many surface locations, which gives a better estimate of the global increase in atmospheric $CO_2$ compared to the single high altitude MLO estimate. (b) Global $CO_2$ bottom-up budget obtained with our estimates of fossil $CO_2$ emissions, DGVM-based net land $CO_2$ flux, net ocean sink from ocean model emulators, and annual $CO_2$ growth rates from MBL. The red curve is -1 * fossil emissions and the difference between the bars and this curve is the imbalance of the bottom-up budget. (c) Global $CO_2$ top-down budget obtained with our estimates of fossil $CO_2$ emissions and inversions assimilating OCO-2 atmospheric $CO_2$ mixing ratios. By design, inversions match OCO-2 data and their growth rate is the blue bar. The orange squares are the $CO_2$ growth rate obtained directly from OCO-2 data using the GRESO model.

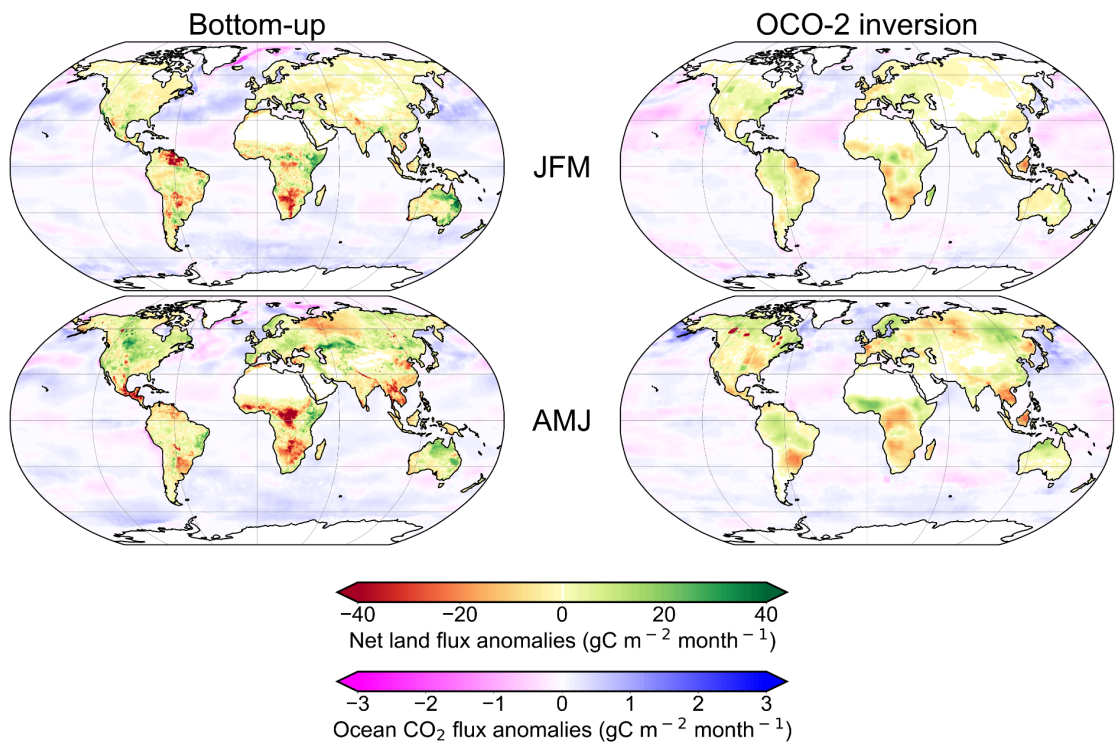

**Fig. 2 Net land and ocean $CO_2$ flux anomalies for January to March (JFM) and April to June (AMJ) 2024.** Anomalies are calculated relative to the 2015-2022 average of the DGVM models (left column, mean of three models) and the OCO-2 inversions (right column, mean of three models). Positive values represent anomalous $CO_2$ uptake from the atmosphere by the land (green) or ocean (blue).

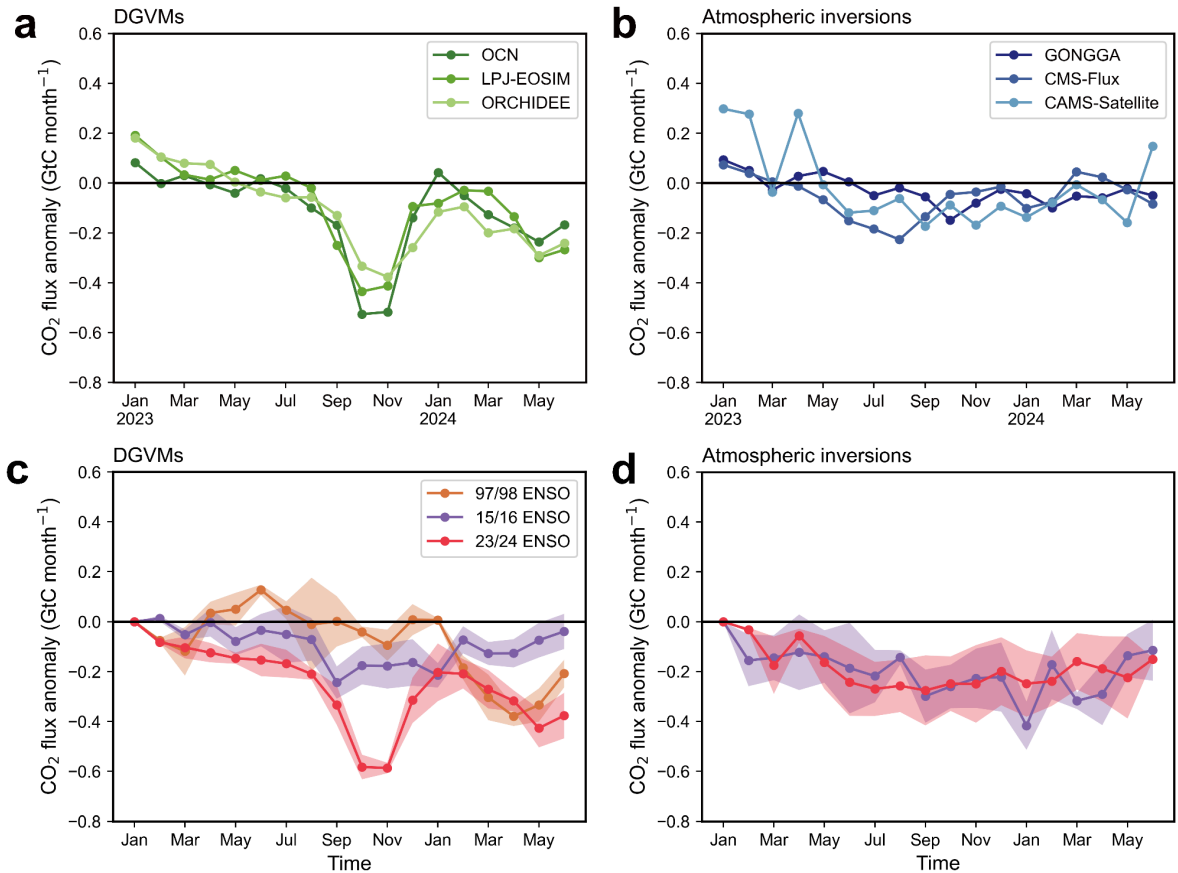

**Fig. 3 Evolution of monthly tropical $CO_2$ flux anomalies.** (a) Anomalies from the three DGVMs used in this study, starting from January 2023, showing a switch to $CO_2$ sources anomalies denoted by negative values starting in September 2023 and lasting until at least July 2024. (b) Same for the three inversions. (c) Comparison of mean $CO_2$ flux anomalies and 1-sigma standard deviation between models (shaded areas) of the three DGVMs for the last three El Niño events (1997/98, 2015/16, and 2023/24). (d) Same for the three inversions which start in 2015 and therefore do not cover the 1997/98 event.

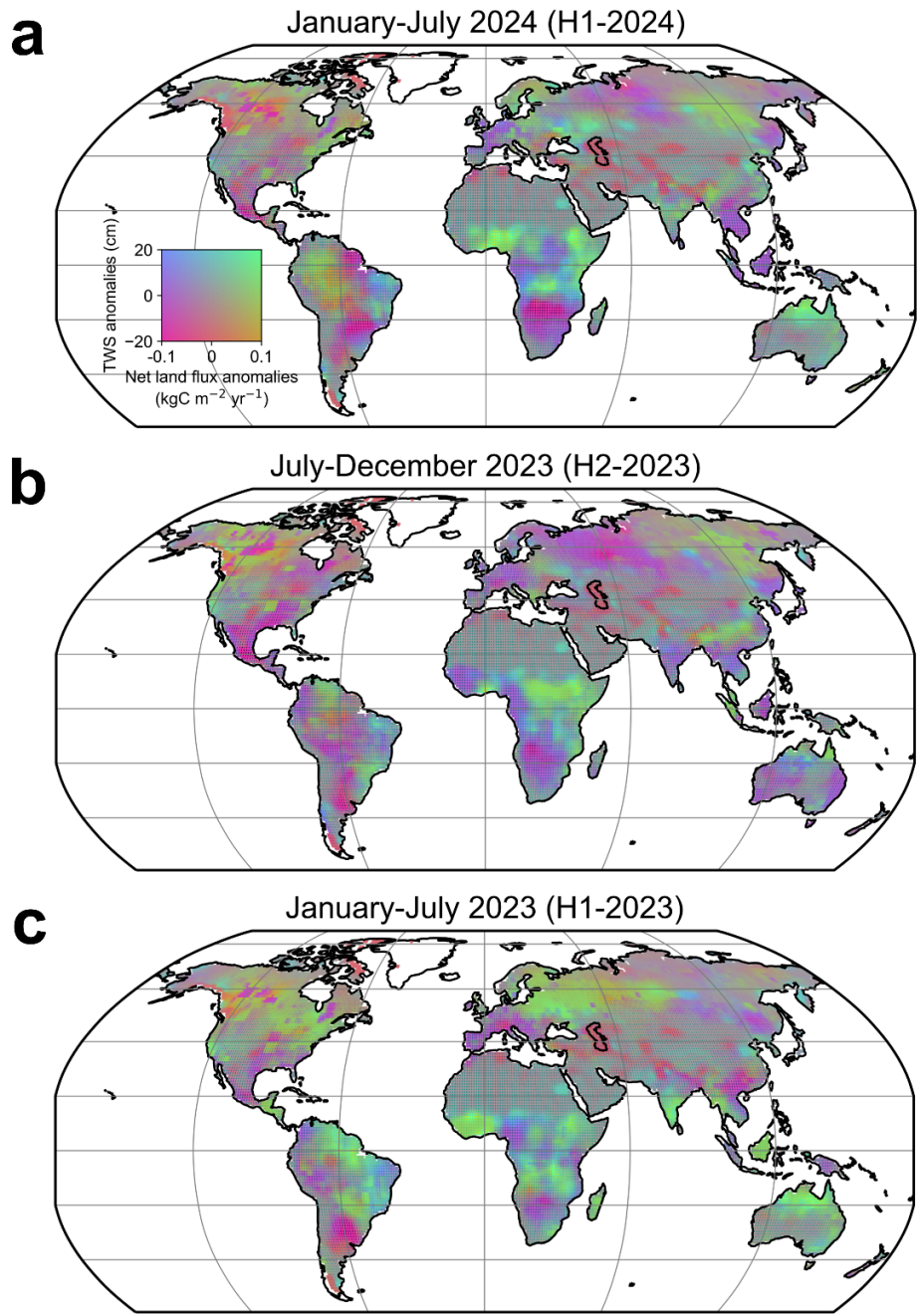

**Fig. 4 Bivariate plots showing co-variations between net land flux anomalies from inversions and total water storage anomalies from the GRACE satellites in H1-2024, H2-2023 and H1-2023.** Green areas show wetter anomalies coincident with more $CO_2$ uptake, and magenta areas show dryer anomalies that are coincident with reduced $CO_2$ uptake.

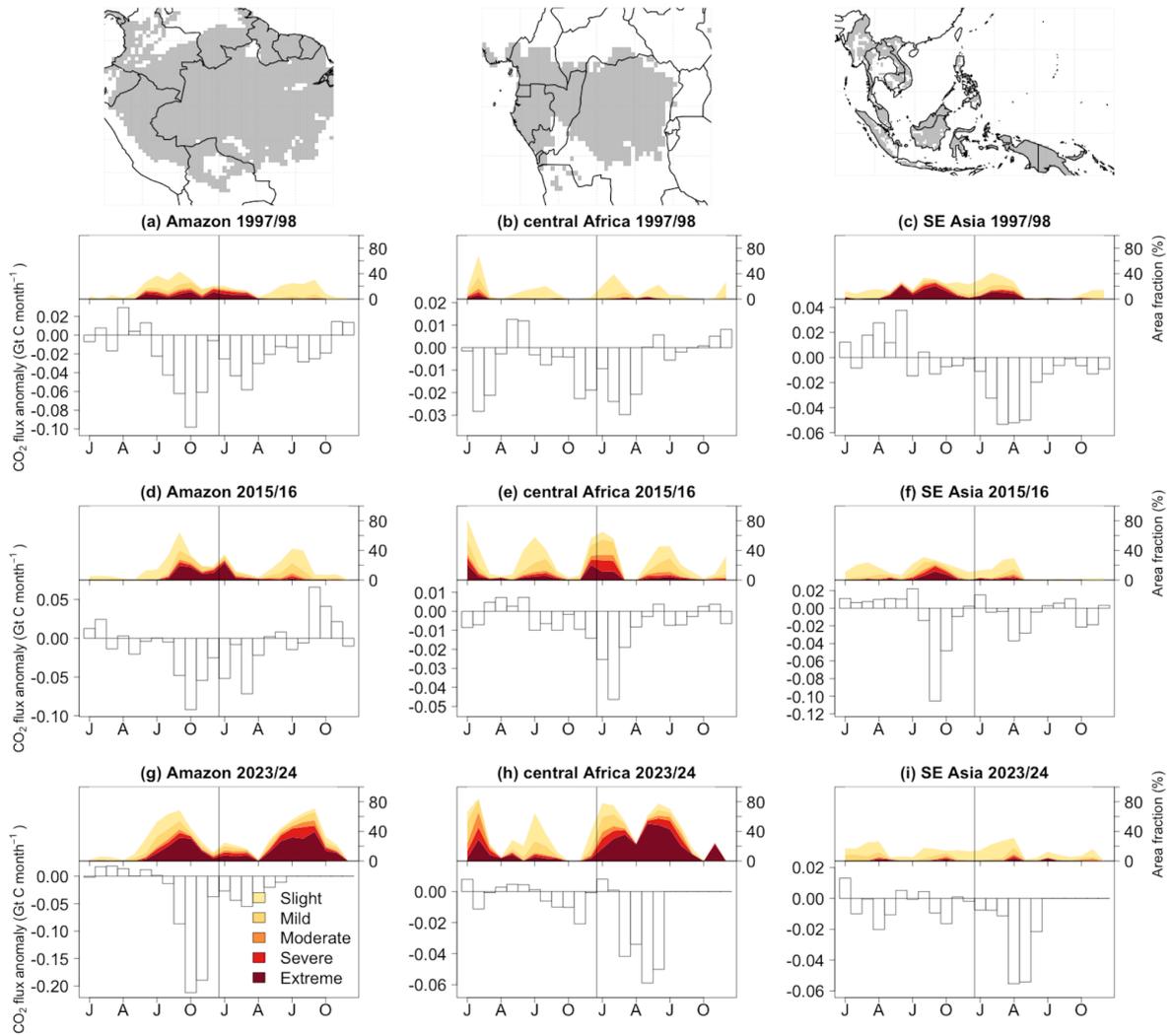

**Fig. 5. Drought timing and extent of the three El Niño events with net CO$_2$ flux anomalies.** Upper part of each subplot displays the fraction of each tropical continental area in grey under drought for different intensities. Fractional area under each drought severity category is calculated for different severities based on Z score transformed monthly cumulative water deficits. Lower part displays monthly net land CO$_2$ fluxes anomalies from the mean of three DGVMs (ORCHIDEE, OCN, LPJ-EOSIM) calculated relative to the 5-years pre-event baseline for each El Niño. **(a-c)** For the 1997/98 El NIño in the Amazon, central Africa and tropical Southeast Asia. **(d-f)** Same for the 2015/16 El Niño. **(g-h)** Same for the 2023/24 El Niño. The anomalies for this El Niño are only simulated until July 2024.

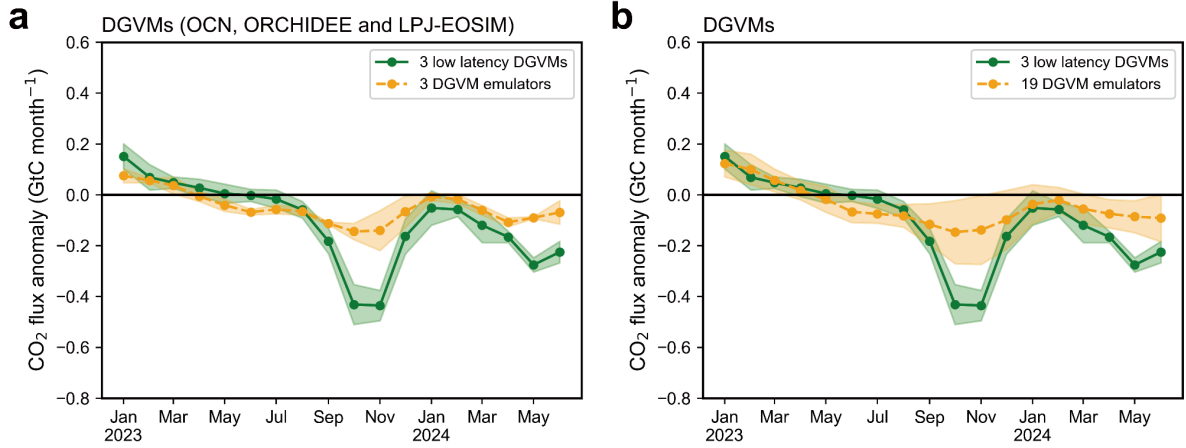

**Fig. 6. Comparison of monthly tropical $CO_2$ flux anomalies from January 2023 to July 2024 between ML emulators of DGVMs and the three DGVMs used in this study.** **(a)** Comparison between ML emulators of the same three DGVMs used in this study (OCN, ORCHIDEE and LPJ-EOSIM) in yellow and their original model simulations in green. The emulators are trained using runs of each DGVM until 2023 provided for the global carbon budget (GCB2024) [9]. The ORCHIDEE and LPJ-wsl models used in GCB2024 are different from the ORCHIDEE-MICT and LPJ-EOSIM versions used in this study. **(b)** Comparison of the three low-latency DGVMs used in this study with the emulators of 19 DGVMs from GCB2024 with monthly outputs.

# Supplementary

1. **Global July-July annual growth rates of atmospheric $CO_2$ from different atmospheric observations and bottom-up models**

We compared the atmospheric $CO_2$ growth rate from July 2010 to July 2024 based on in-situ observations from 40 marine boundary layer background stations, calculated by NOAA GML[1,2] as the year on year difference between smoothed observations between May-August averaged across all the stations (blue bars), from the Mauna Loa station (dark blue squares), from the assimilation of global OCO-2 satellite observations of column-average $CO_2$ dry air mole fraction retrievals, about 300,000 10-second-averaged retrievals each year, by the inversions used in this study (brown curve), OCO-2 satellite data processed with the data-driven Growth Rate From Satellite Observations (GRESO) approach of ref. [5] (orange squares), and by the bottom-up approach, that is, not using atmospheric measurements and the growth rate is predicted from the difference between fossil $CO_2$ emissions minus the land sink from three DGVM models, minus the ocean sink from ocean model emulators (red curve) (Supplementary Fig. 1c).

The Growth Rate From Satellite Observations (GRESO) method [5] processes OCO-2 satellite soundings to derive $CO_2$ growth rates. It involves two main stages: (a) sequential aggregation and (b) time series processing. First, individual soundings are aggregated into weighted 10-second averages, filtered for quality and outliers, and then gridded (5°x5°, 16-day). These gridded data are averaged spatially (first by longitude, then using latitude-weighting) to create a representative global $CO_2$ time series. Second, the $CO_2$ time series is interpolated, deseasonalized, and sampled at the edge of months. Then, consecutive month-edge values are different to calculate deseasonalized monthly and, subsequently, annual $CO_2$ growth rates. Additional 5-month smoothing is applied on monthly growth rates to reduce noise. Here we used only the LNLG (Land Nadir Land Glint) OCO-2 $XCO_2$ data. The GRESO method is described in detail in [5].

2. **Maps of monthly air-sea $CO_2$ fluxes from emulators of biogeochemical and data driven ocean models**

We utilized a data-driven deep learning technique for near-real-time estimation of oceanic carbon monthly gridded fluxes [53]. This method integrates year, month, latitude, longitude, and nine environmental factors as predictors, targeting predictions for each GOBM model or ocean data product. We utilize monthly data up to the end of 2021 from 5 GOBMs and 8 data products in the Global Carbon Budget 2022, with $fCO_2$ output provided at a 1° × 1° resolution. Our predictive variables include a range of biological, chemical, and physical factors typically linked to fluctuations in $fCO_2$. These variables are sea surface temperature (SST), sea ice fraction (ICE), sea surface salinity (SSS), atmospheric $CO_2$ mole fraction ($xCO_2$), mixed-layer depth (MLD), sea surface height (SSH), chlorophyll a (chl a), sea level pressure (SLP), and wind speed. All data are bilinearly interpolated to a 1° × 1° monthly resolution to align with our $fCO_2$ targets and updated to June 2024. Since the $xCO_2$ data is only available up to the end of 2023, and to meet the requirement for near-real-time data, we gather global average marine boundary layer surface monthly mean atmospheric $CO_2$ data updated to June 2024. We use a Linear Regression model (LR) model to establish a relationship among the year, month, latitude, longitude, mean atmospheric $CO_2$ data, and $xCO_2$. Additionally, we incorporated valuable features: $xCO_2$ data from the same month over the past six years. We used data from 2015-2022, and testing on 2023 data yielded a test RMSE of 0.40, reflecting roughly a 0.1% prediction error. This approach allows us to extend the $xCO_2$ data to near-real-time. All data are formatted into a 180x360 grid and subdivided into 18x18 patches for computational efficiency. During the training process, we incorporated predictors from the current month, as well as predictors and historical $fCO_2$ data from the past 12 months, as auxiliary information. Consequently, the input data is ultimately processed into a size of 18x18x14. Our experiments demonstrated that including auxiliary information from the previous 12 months significantly enhances the model's accuracy. This improvement is likely due to the lagged effects of some predictors and the corrective influence of historical $fCO_2$ data on the model's predictions.

After processing the data, we employed a Semi-supervised learning framework combined with Convolutional Neural Network (CNN) models for prediction. Our architecture integrates multi-layered CNNs and linear models to efficiently process the input data. The CNNs, inspired by human visual perception, consist of Convolutional Layers, Rectified Linear Unit (ReLU) Layers, and Fully Connected Layers, which collectively transform input data into predictions. The input layer has dimensions of 18x18x14, followed by two hidden layers of 18x18x64 that learn spatial hierarchies. Linear layers, also 18x18x64, perform linear transformations. The output layer, with dimensions of 18x18x1, outputs the predicted oceanic carbon $fCO_2$ value. To enhance stability in areas with missing values and at the land-sea interface, we used a semi-supervised model framework. Initially, the CNN was trained on labeled data, calculating the Root Mean Square Error (RMSE) between labels and predictions as the supervised loss (LsL_sLs). For prediction stability on unlabeled data, we used pseudo-labeling: predicting with 10% of input predictors removed as pseudo-labels, then predicting again with 30% of input predictors removed, calculating the RMSE as the unsupervised loss (LuL_uLu). The model was updated using the weighted sum of LsL_sLs and LuL_uLu through backward propagation. This architecture enables efficient and accurate prediction of oceanic carbon flux.

## 3. Maps of monthly terrestrial CO$_2$ fluxes from emulators of Dynamic Global Vegetation Models

We applied a deep-learning emulator framework identical to that used for oceanic fluxes to estimate monthly terrestrial carbon fluxes. Specifically, we utilized monthly terrestrial CO$_2$ flux data from simulation 2 of 19 Dynamic Global Vegetation Models (DGVMs) included in the Global Carbon Budget 2024 [9], covering the period up to December 2023 at a standardized spatial resolution of 0.5° × 0.5°. Predictor variables comprised atmospheric CO$_2$ mole fraction and eight meteorological variables from the ERA5 [22]: surface pressure, surface downward short-wave radiation, surface downward long-wave radiation, 2-meter temperature, total precipitation, 10-meter wind components, and specific humidity. All predictors were bilinearly interpolated to the uniform monthly spatial resolution. The model was trained and validated using data spanning the period from 2000 to 2023, and subsequently employed to predict H1-2024. Additionally, we conducted performance testing on the emulators. Specifically, data from 2022 to 2023 was withheld from the training and validation datasets (2000–2021) to serve as an independent test set for evaluating model performance. Supplementary Fig. 3 provides a detailed comparison between emulator predictions and their corresponding DGVM outputs (their low-latency versions were used in this study). The results demonstrate excellent agreement, highlighting the robustness and accuracy of the emulator framework.

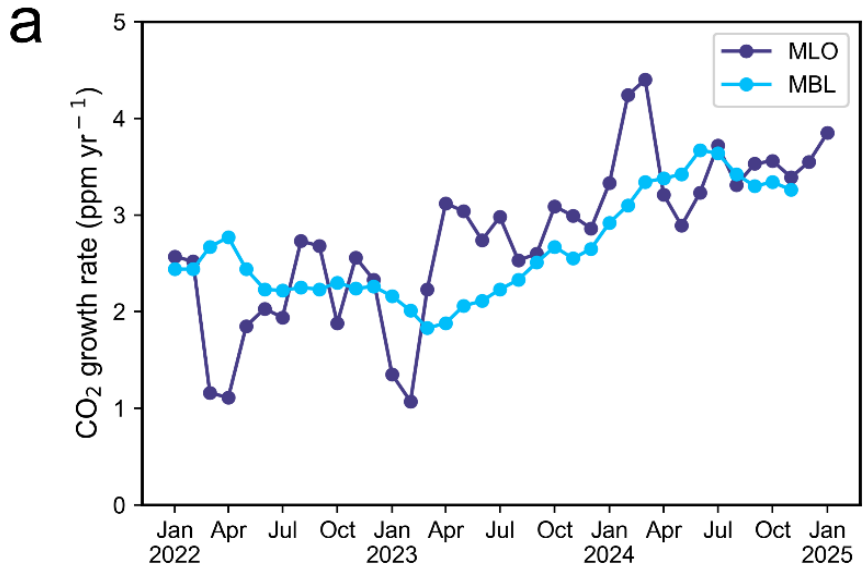
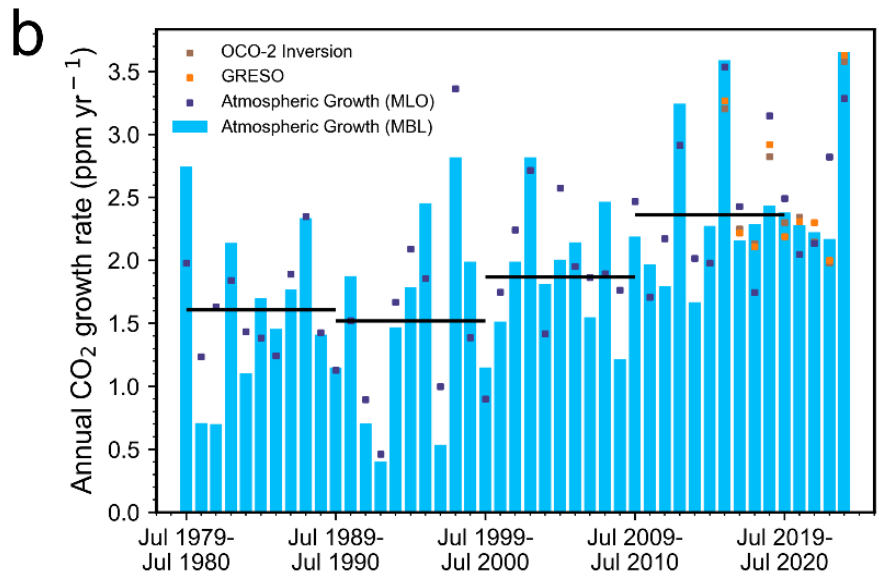
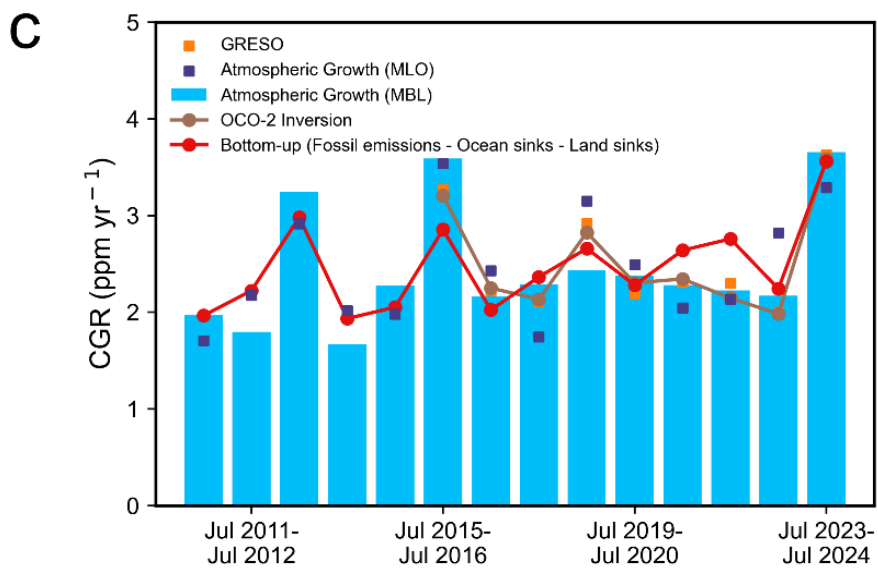

**Supplementary Figure 1 (a)** Monthly running mean of annual $CO_2$ growth rates with marine boundary layer surface stations in blue (MBL) and Mauna Loa (MLO) in dark blue (MLO) from January 2022 to January 2025. **(b)** Annual June-July to June-July growth rates from MBL stations (blue bars), MLO (dark blue), OCO-2 column $CO_2$ data assimilated by inversions (brown squares), and OCO-2 satellite data processed with the data-driven GRESO approach of ref. [5] (orange squares). **(c)** June-July to June-July $CO_2$ growth rates from July 2010 to July 2024 from different approaches, including MBL stations (blue bars), MLO (dark blue squares), the assimilation of global OCO-2 column $CO_2$ concentration (brown curve), GRESO (orange squares), and the bottom-up approach where the growth rate is calculated from the difference between fossil fuel $CO_2$ emissions minus the sum of the net land $CO_2$ flux from three DGVM models and the ocean sink from ocean model emulators (red curve).

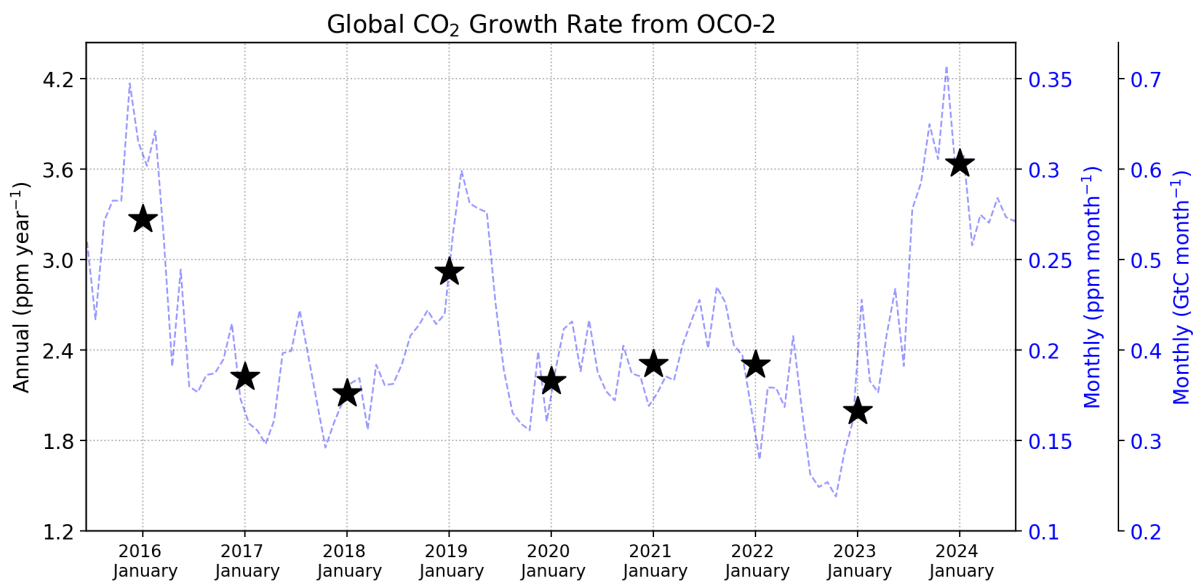

**Supplementary Figure 2 Annual and monthly global $CO_2$ growth rates from OCO-2 data using the data-driven GRESO approach.** Black stars indicate the annual growth rate measured from June–July to June–July of consecutive years (left y-axis, in ppm year$^{-1}$), while the blue dashed line represents the monthly growth rate (from the start to the end of each month; right y-axis, in ppm/month). A five-month running mean is applied to the monthly time series to reduce noise. Growth rate values were calculated using the Growth Rates Using Satellite Observations (GRESO) approach from ref. [5].

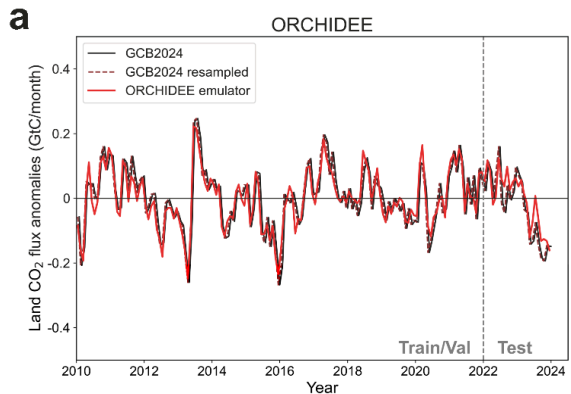
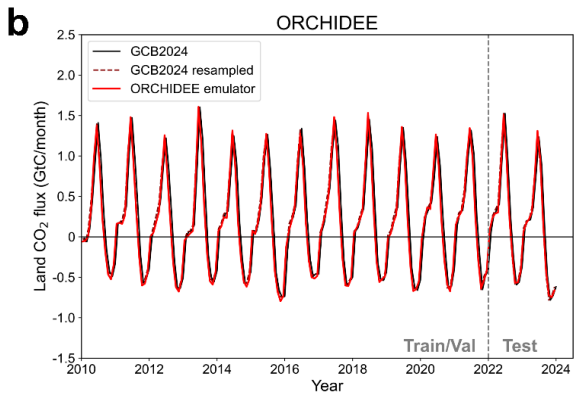
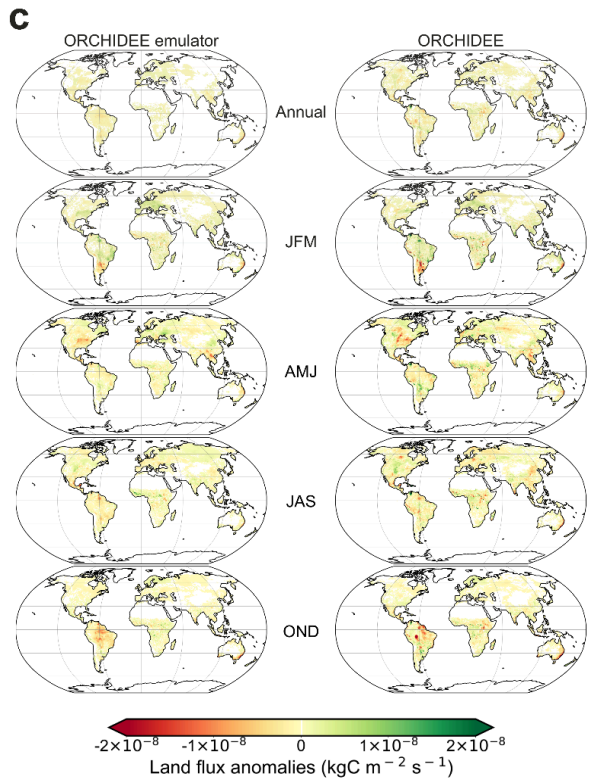
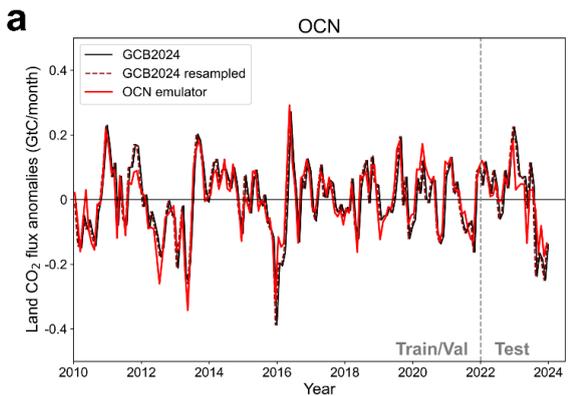
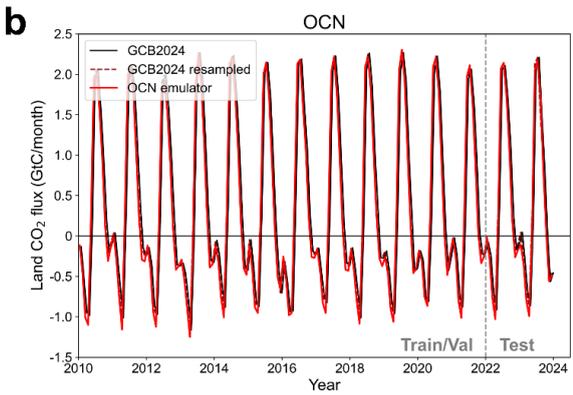
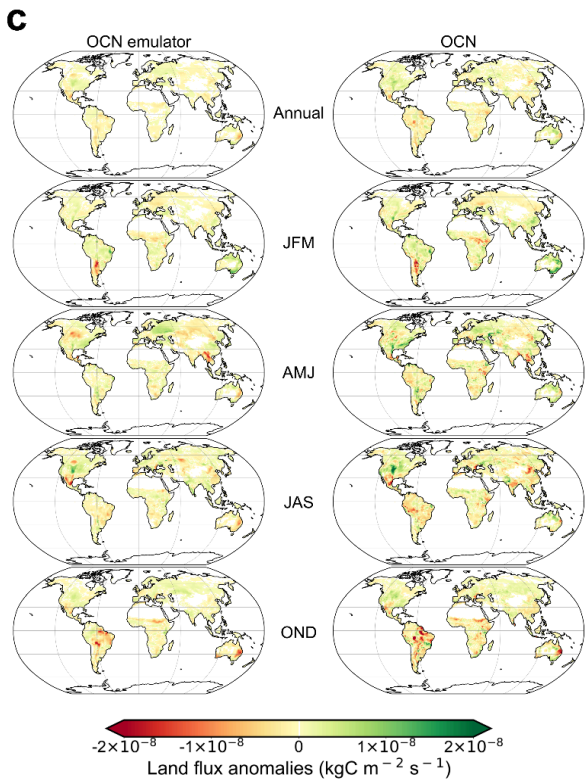

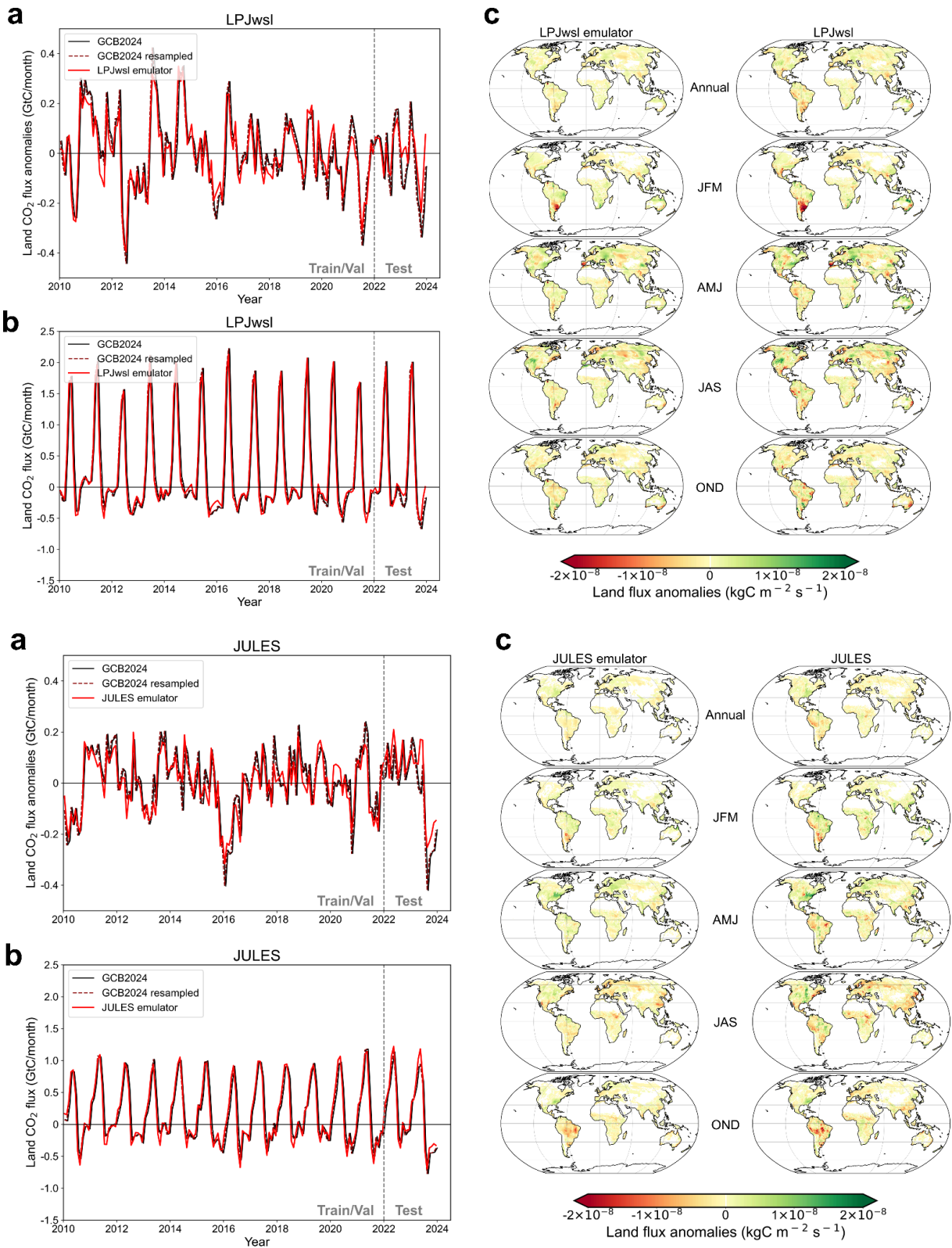

**Supplementary Figure 3. Comparison of DGVM emulators with their original models for ORCHIDEE, OCN, LPJ-wsl, and JULES.** The emulators were trained and validated using data from 2000–2021, and tested in 2022 and 2023. **(a)** Time series of deseasonalized monthly anomalies from 2010-2023 with original models in black and emulator results in red. The red dashed line represents the original model outputs resampled to 0.5° resolution, which served as the training target for all emulators (top left). **(b)** Time series of net land $CO_2$ fluxes from Jan. 1st 2010 to Dec. 31st 2023

(bottom left), and **(c)** spatial distribution of net land $CO_2$ fluxes anomalies for each quarter in 2023 with emulator on the left and native model on the right.

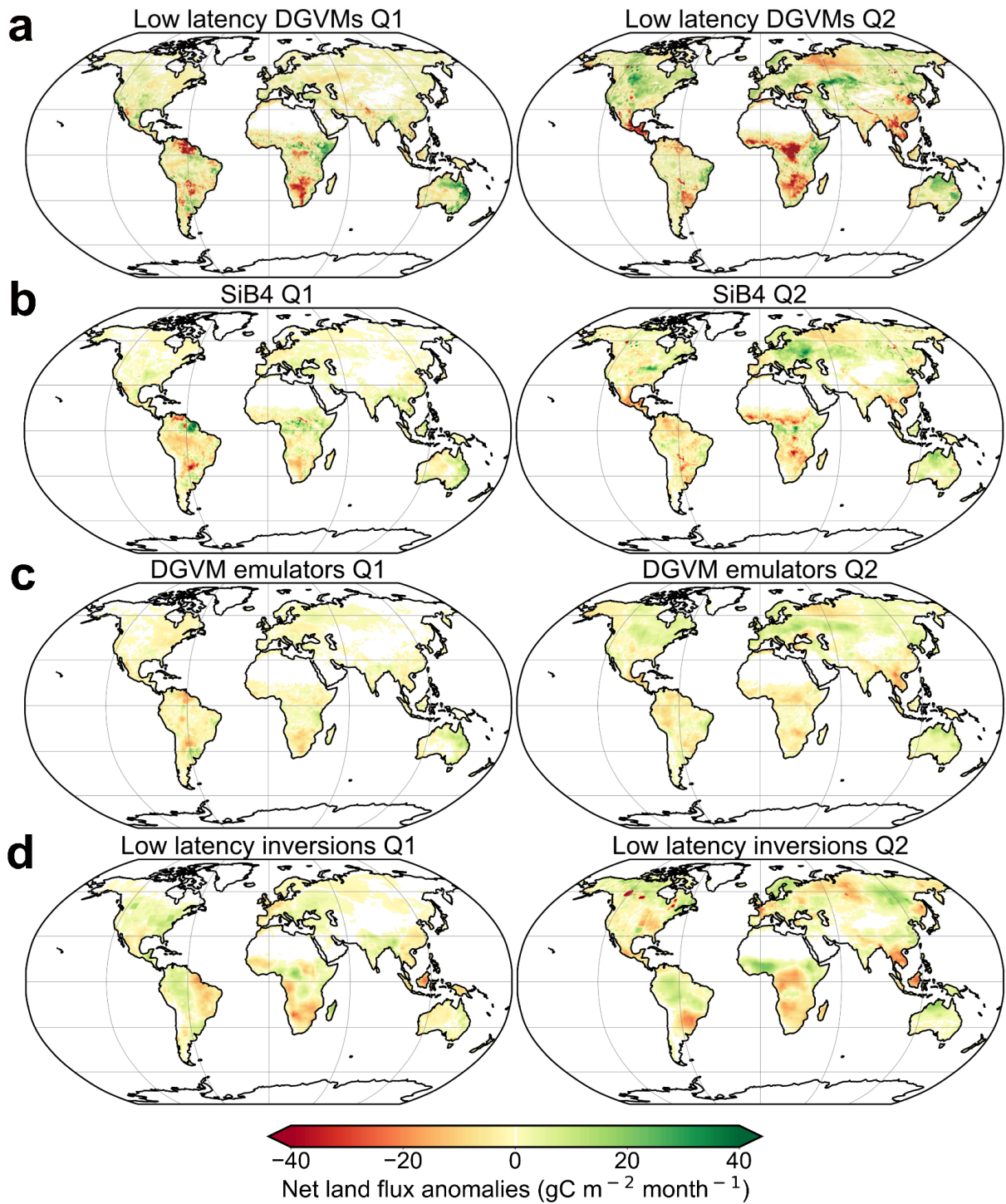

**Supplementary Figure 4. Net land $CO_2$ flux anomalies for the first and second quarters in 2024 (Q1 = JFM and Q2 = AMJ).** Anomalies are calculated from the 2015-2022 average of each quarter. **(a)** mean of the three low latency DGVMs used in this study, **(b)** SiB4 flux model, **(c)** mean of the 19 DGVM emulators and **(d)** mean of the three OCO-2 inversions. Positive values in green represent increased flux from the atmosphere to the land, that is anomalous carbon sinks.

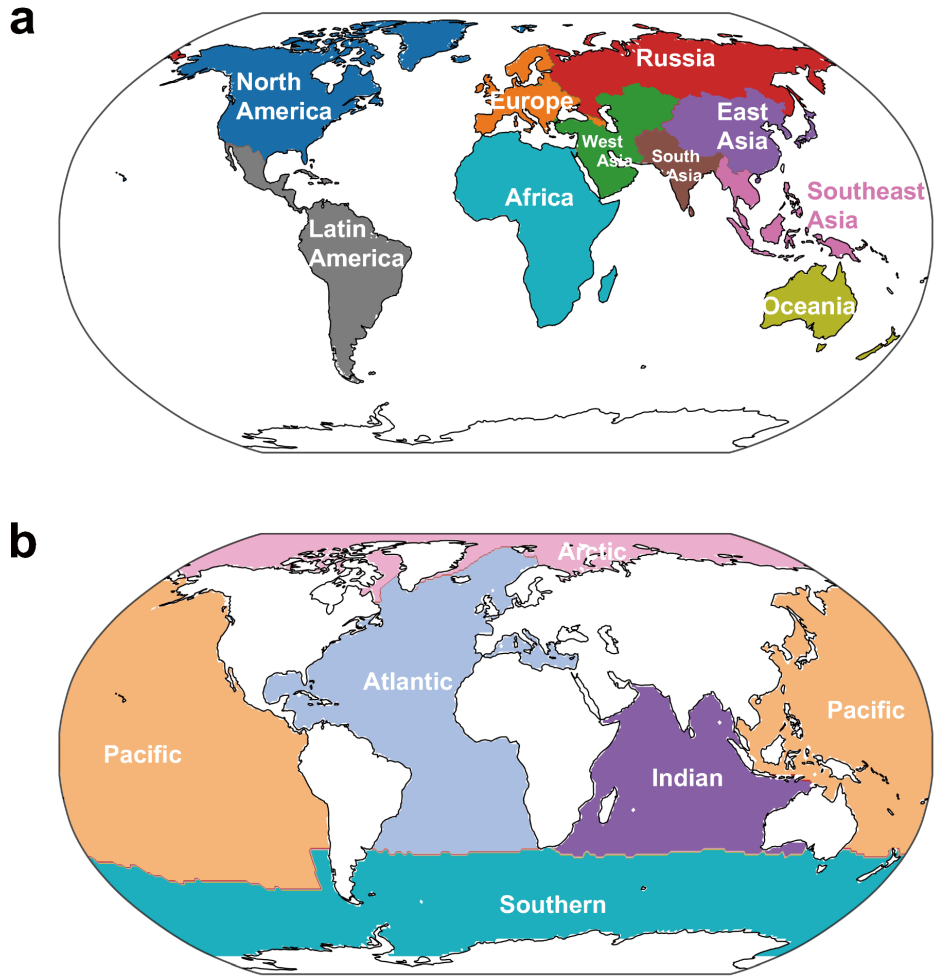

**Supplementary Figure 5. Maps of RECCAP2 land (a) and ocean (b) regions.**

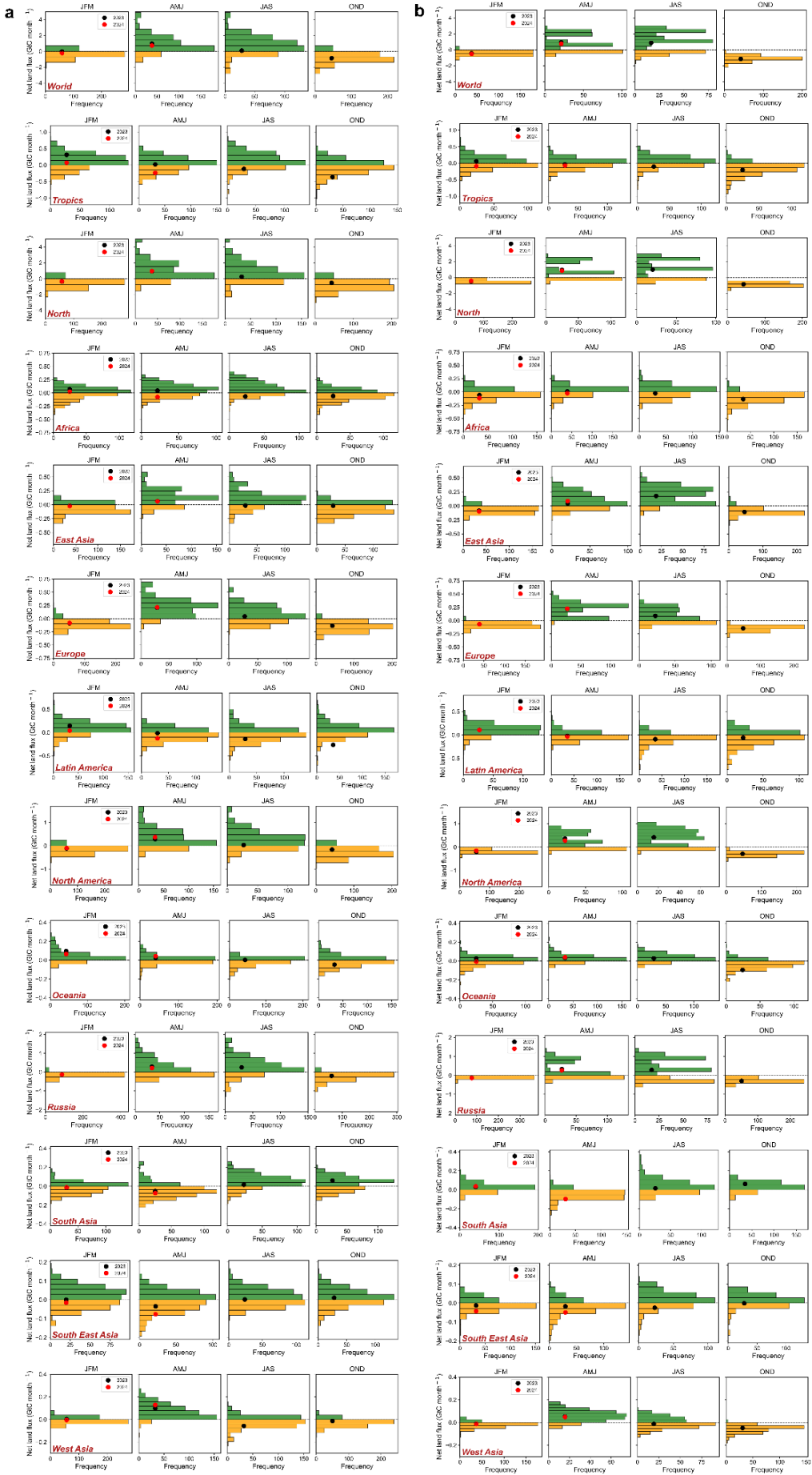
36

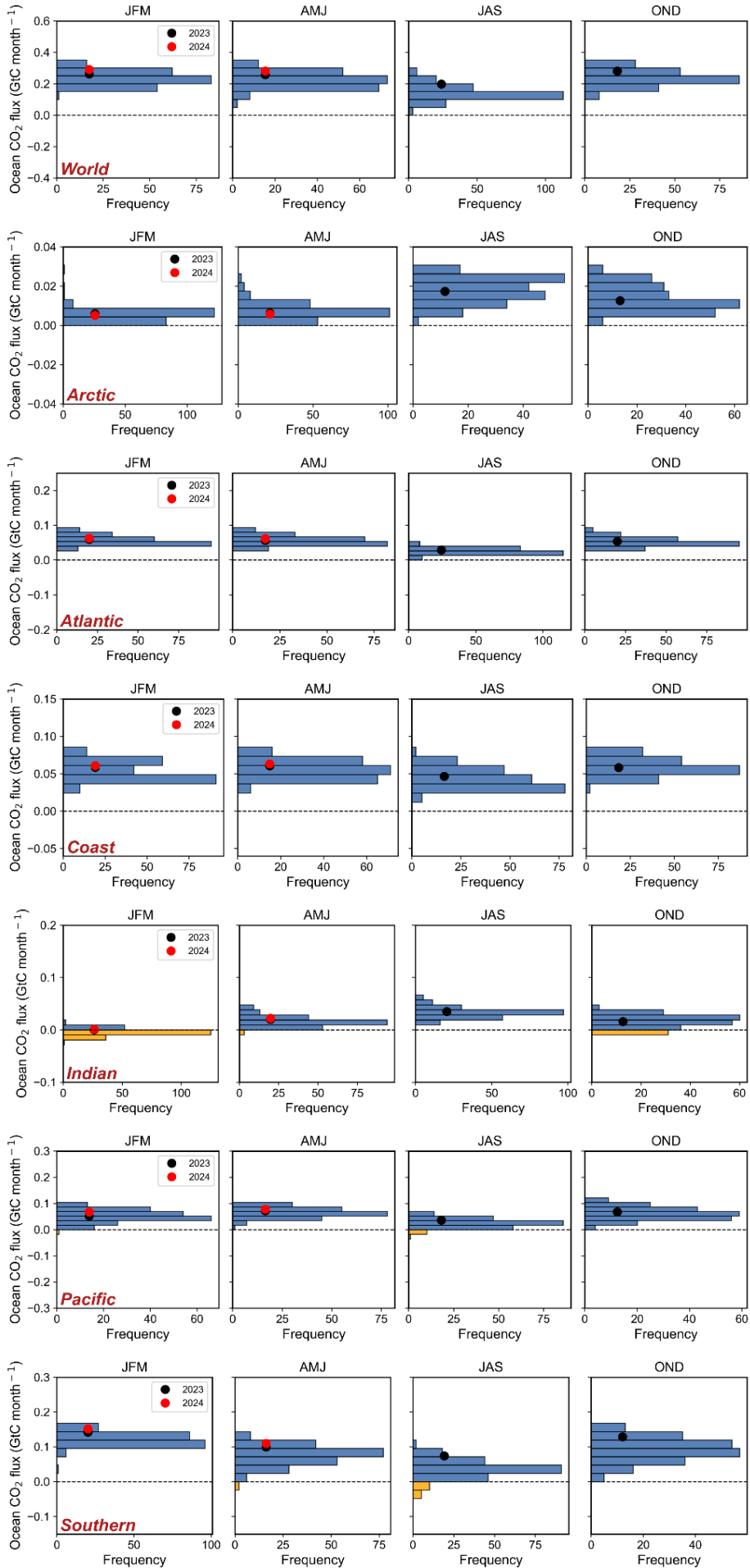

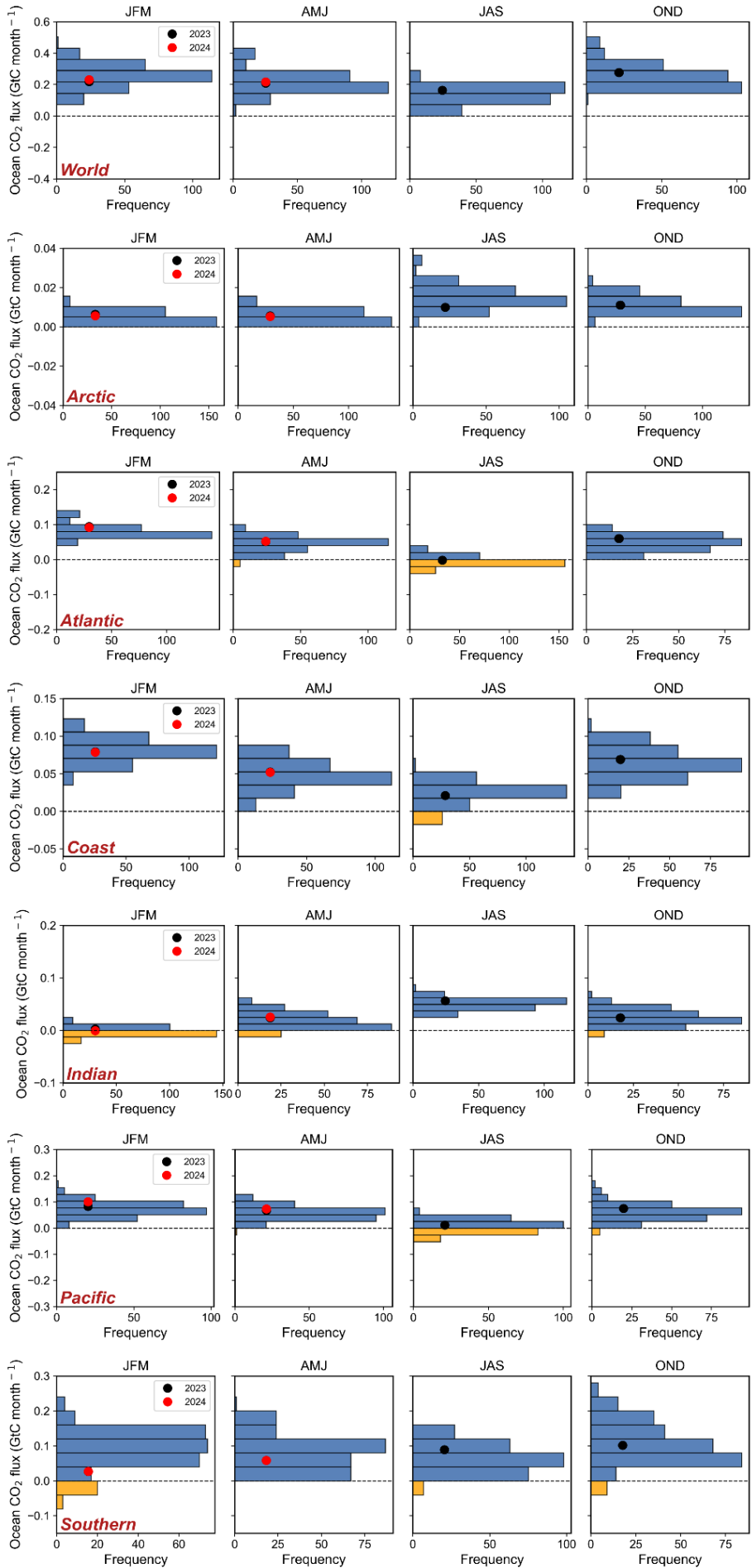

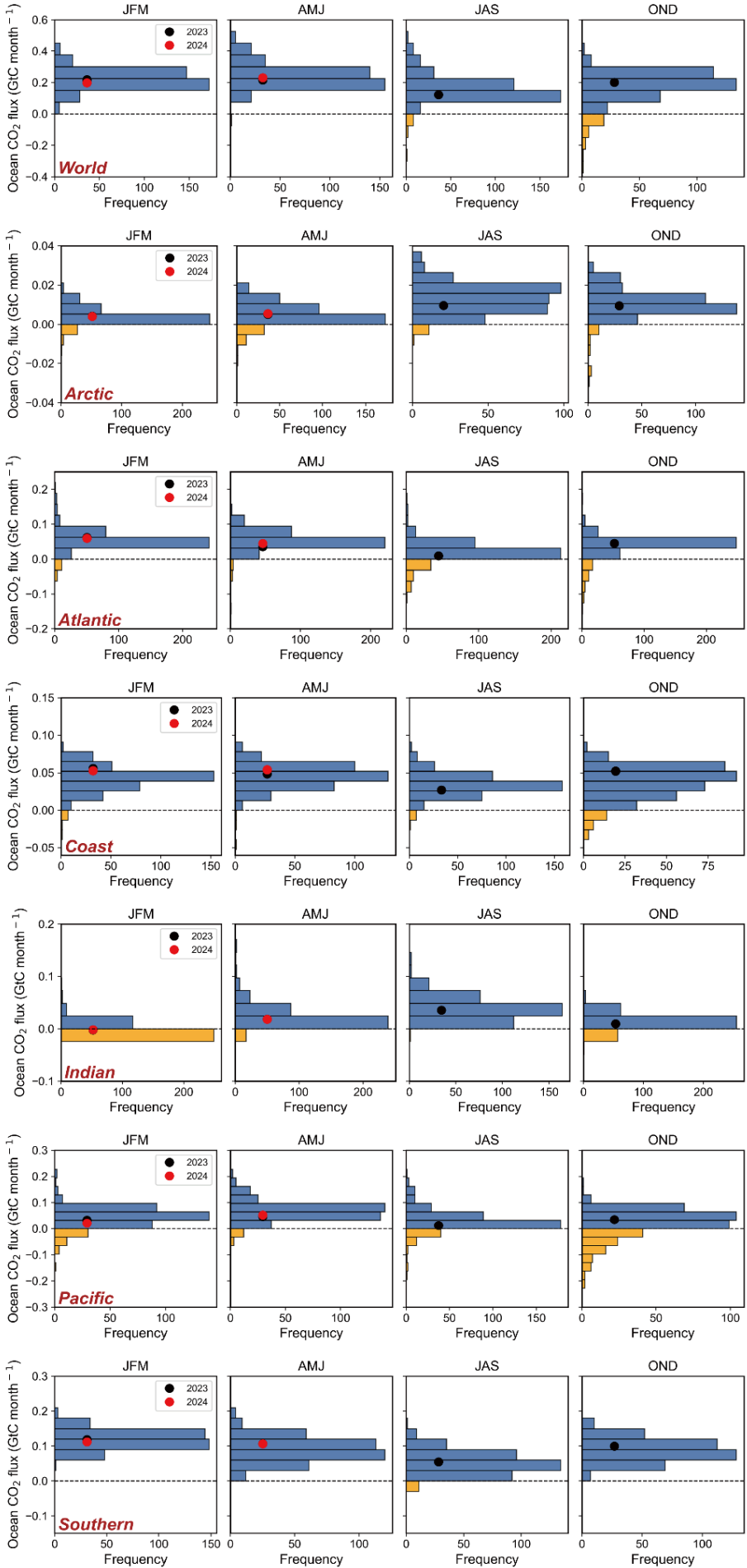

**Supplementary Figure 6. Quarterly land and ocean fluxes for the RECCAP2 regions in 2023 (black dots) and in the first six months of 2024 (red dots). Fluxes are shown with the distributions of these fluxes during 2015-2023 with blue / green bars being positive values (sinks) and yellow ones negative values (sources). (a)** Quarterly land flux for each RECCAP2 land region in the distribution of the fluxes from the DGVM models. **(b)** Quarterly land flux from the OCO-2 inversions for each RECCAP2 land region in the distribution of the flux from the inversion models. **(c)** Quarterly ocean flux from the ocean model emulators trained on GCB2024 ocean data-driven flux products for each RECCAP2 ocean region in the distribution of the flux from the data products models. (d) Quarterly ocean flux from the ocean model emulators trained on GCB2024 Global Ocean Biogeochemical Models (GOBM) for each RECCAP2 ocean region, in the distribution of the flux from the GOBMs. **(e)** Quarterly ocean fluxes from the OCO-2 inversions for each RECCAP2 ocean region in the distribution of the flux from the inversion models. See Supplementary Figure 5 for the maps of the RECCAP2 regions.

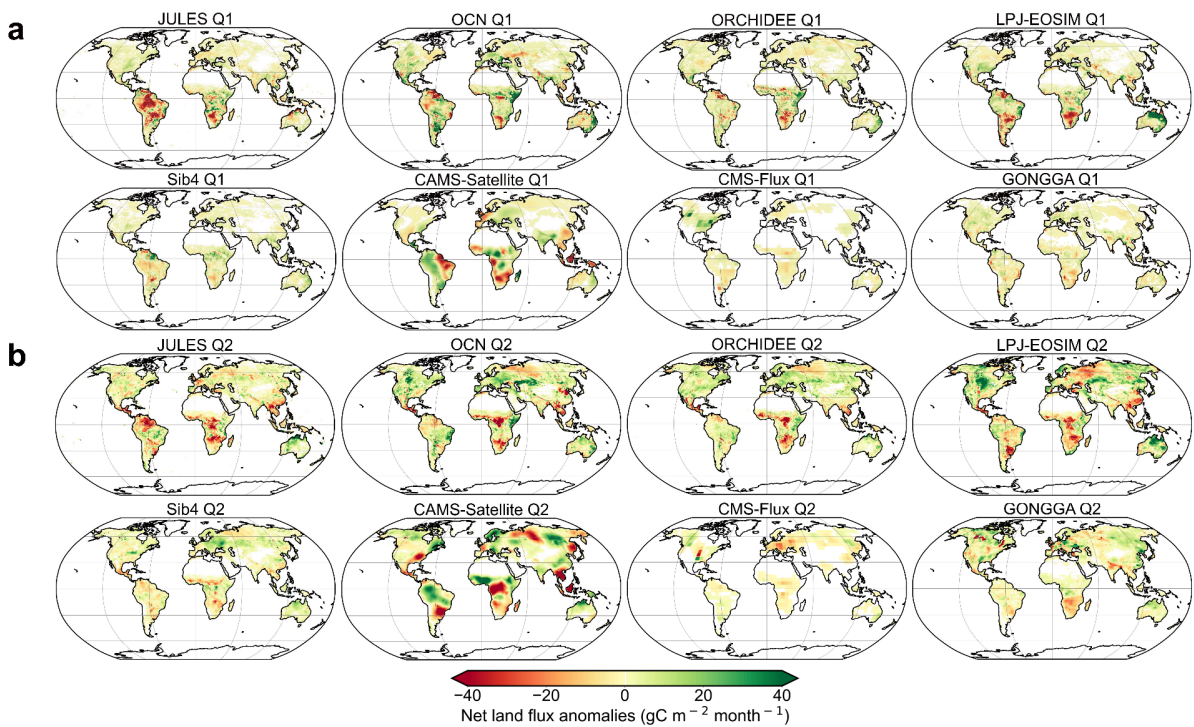

**Supplementary Figure 7. Net land CO$_2$ flux anomalies for the first and second quarters in 2024 from each model.** The top two rows show fluxes during JFM 2024 (Q1) and the bottom two during AMJ 2024 (Q2).

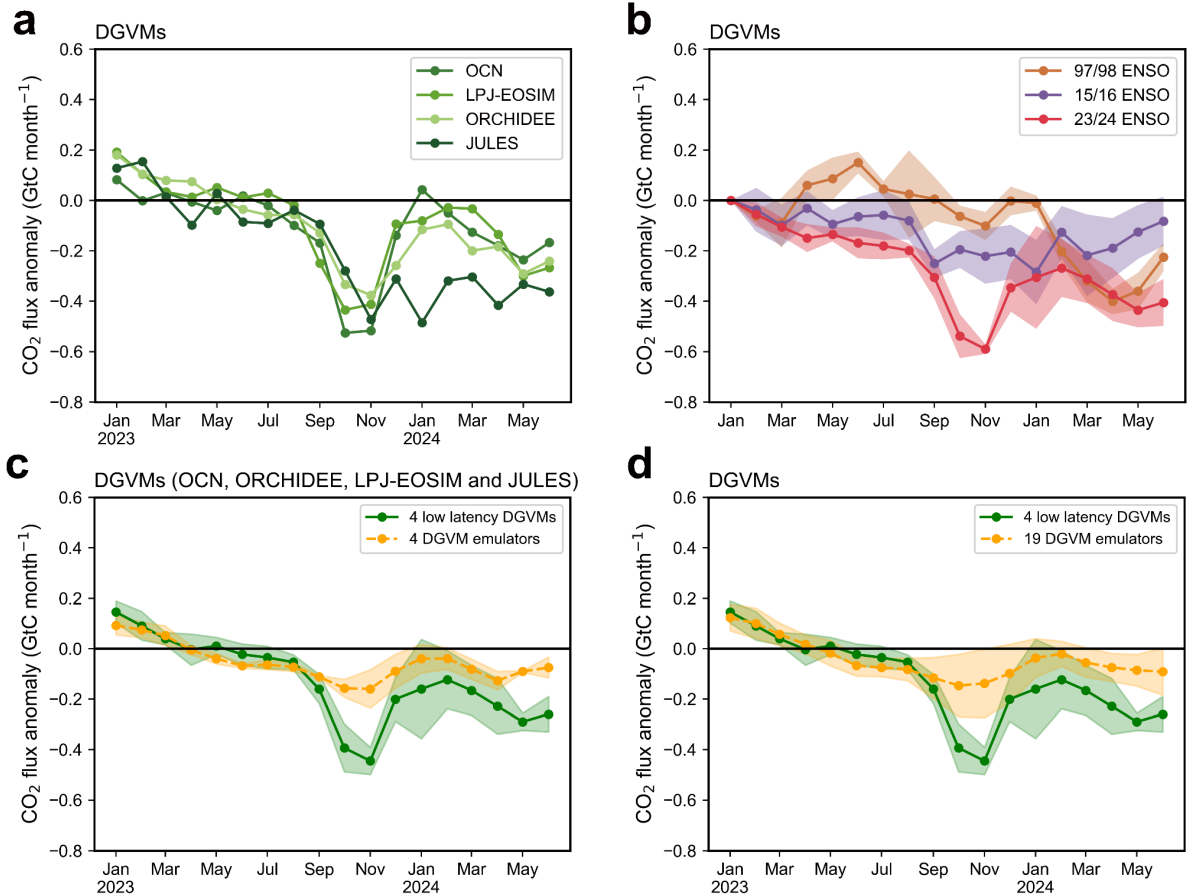

**Supplementary Figure 8. Evolution of monthly tropical CO$_2$ flux anomalies. (a)** For the three DGVMs used in this study included JULES which was not used in the assessment because of its extremely low CO$_2$ source anomaly in 2023 and 2024. **(b)** Comparison of mean CO$_2$ flux anomalies and 1-sigma standard deviation between models (shaded areas) of the three DGVMs used in this study plus JULES for the last three El Niño events. **(c)** Comparison between the mean of ML emulators of four DGVMs ( OCN, ORCHIDEE and LPJ-EOSIM used in this study plus JULES) (yellow) and their original model simulations (green). The emulators are trained using runs of each native DGVM until 2023 from GCB2024 [9]. The ORCHIDEE and LPJ-wsl models versions used in GCB2024 are different from the ORCHIDEE-MICT and LPJ-EOSIM versions used in this study. **(d)** Comparison of the mean of the emulators of 19 DGVMs (yellow) with the four DGVMs including JULES (green).

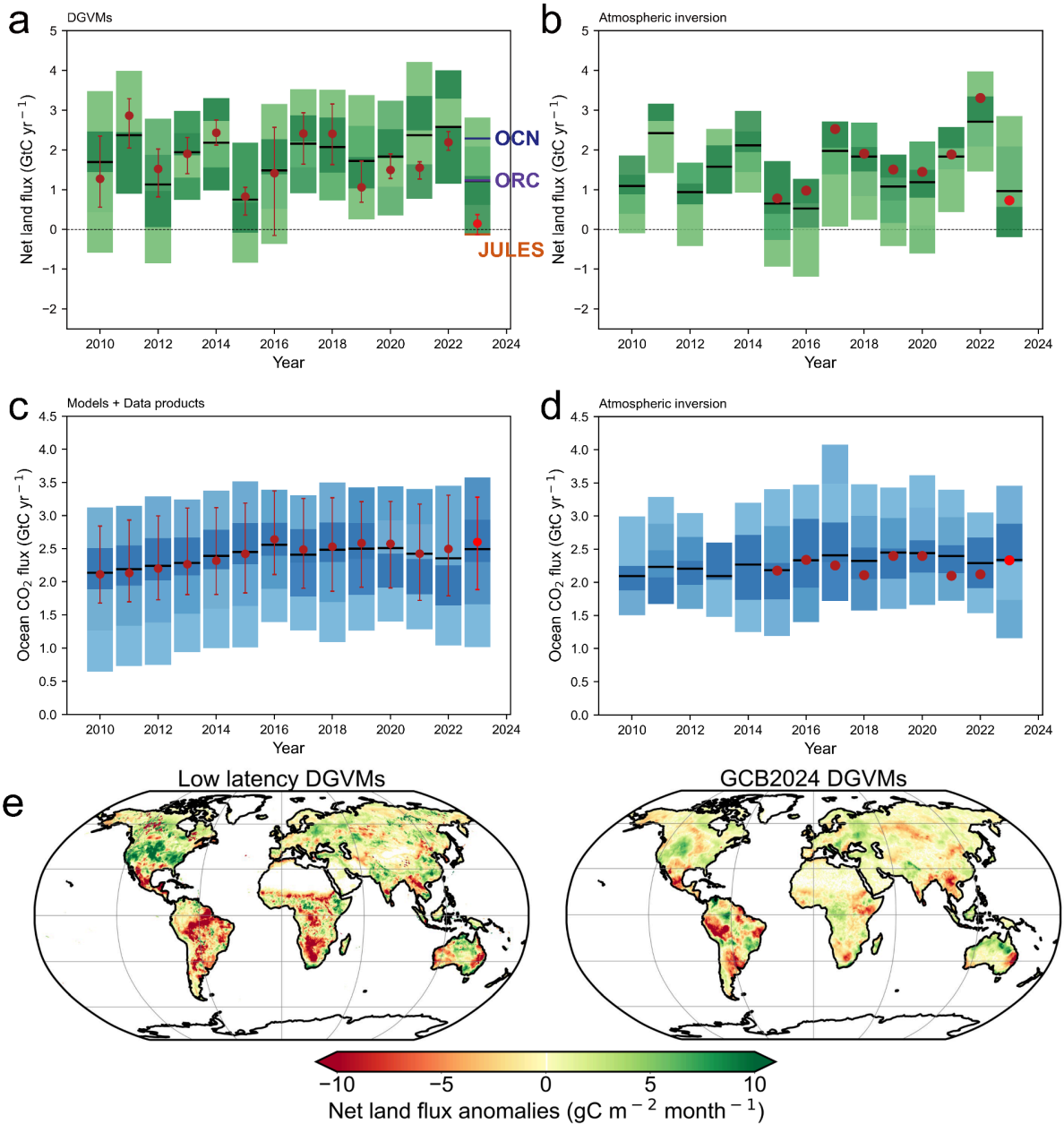

**Supplementary Figure 9 Low latency budgets 2023 compared with the GCB 2024 model results.**
**(a-d)** The red dot shows the mean of the low latency models used in Ref. [3] and lines indicate the position of each DGVM model included in Ref. [3], as derived from the GCB24 results. Green bars (blue for ocean fluxes) indicate the range of the GCB2024 models with darker colors indicating that more models are in this interval. **(e)** Net land $CO_2$ flux anomalies in 2023 from the 2015-2022 period for the low latency DGVM models used by Ref. [3] for the $CO_2$ budget of the year 2023 and the mean of the TRENDY DGVMs used in GCB2024. Positive values (in green) represent increased flux from the atmosphere to the land or ocean (anomalous carbon sink).

**Supplementary Table 1** CO$_2$ fluxes (2015–2022 average, 2023 and H1-2024) for each land RECCAP2 region from DGVMs and inversion methods (2015–2022 from GCB 2024; 2023-2024 from NRT methods; Unit: GtC yr$^{-1}$).

| Period | Methods | World | Tropics | North | Africa | East Asia | Europe | Latin America | North America | Oceania | Russia | South Asia | Southeast Asia | West Asia |
|---|---|---|---|---|---|---|---|---|---|---|---|---|---|---|
| 2015-2022 average | DGVMs | 1.94 ± 0.16 | 0.67 ± 0.11 | 1.27 ± 0.11 | 0.31 ± 0.04 | 0.26 ± 0.03 | 0.10 ± 0.02 | 0.17 ± 0.06 | 0.36 ± 0.03 | 0.05 ± 0.01 | 0.36 ± 0.04 | 0.06 ± 0.02 | 0.08 ± 0.03 | 0.04 ± 0.01 |
| | Inversions | 1.47 ± 0.16 | −0.39 ± 0.11 | 1.80 ± 0.14 | −0.38 ± 0.08 | 0.39 ± 0.11 | 0.18 ± 0.05 | −0.14 ± 0.12 | 0.58 ± 0.09 | 0.08 ± 0.04 | 0.56 ± 0.07 | 0.13 ± 0.06 | 0.01 ± 0.06 | −0.05 ± 0.04 |
| 2023 | DGVMs (including JULES) | 0.45 ± 0.13 | -0.44 ± 0.17 | 0.89 ± 0.21 | -0.02 ± 0.14 | 0.06 ± 0.06 | 0.09 ± 0.09 | -0.83 ± 0.07 | 0.13 ± 0.29 | 0.23 ± 0.13 | 0.65 ± 0.16 | 0.03 ± 0.03 | -0.07 ± 0.07 | 0.09 ± 0.10 |
| | DGVMs (excluding JULES) | 0.58 ± 0.05 | -0.43 ± 0.24 | 1.02 ± 0.23 | 0.09 ± 0.11 | 0.06 ± 0.07 | 0.12 ± 0.12 | -0.84 ± 0.09 | 0.14 ± 0.39 | 0.23 ± 0.17 | 0.68 ± 0.18 | 0.04 ± 0.05 | -0.09 ± 0.09 | 0.13 ± 0.11 |
| | Inversions | 0.76 ± 0.59 | -0.86 ± 0.61 | 1.63 ± 0.16 | -0.62 ± 0.36 | 0.10 ± 0.13 | 0.28 ± 0.14 | -0.17 ± 0.24 | 0.87 ± 0.67 | -0.03 ± 0.14 | 0.61 ± 0.14 | 0.03 ± 0.12 | -0.16 ± 0.06 | -0.06 ± 0.14 |
| H1-2024 | DGVMs (including JULES) | 2.95 ± 2.06 | -0.97 ± 0.47 | 3.85 ± 2.04 | -0.31 ± 0.06 | 0.24 ± 0.47 | 0.80 ± 0.34 | -0.52 ± 0.58 | 1.63 ± 1.39 | 0.65 ± 0.26 | 0.57 ± 1.57 | -0.50 ± 0.51 | -0.56 ± 0.26 | 0.75 ± 0.22 |
| | DGVMs (excluding JULES) | 2.84 ± 2.91 | -0.56 ± 0.32 | 3.26 ± 2.76 | -0.31 ± 0.09 | 0.13 ± 0.61 | 0.78 ± 0.41 | -0.08 ± 0.44 | 1.52 ± 1.65 | 0.67 ± 0.42 | 0.40 ± 1.58 | -0.60 ± 0.63 | -0.61 ± 0.30 | 0.87 ± 0.31 |
| | Inversions | 1.42 ± 1.04 | -0.97 ± 0.08 | 2.17 ± 1.02 | -0.83 ± 0.73 | -0.06 ± 0.30 | 0.94 ± 0.52 | 0.45 ± 1.27 | 0.72 ± 0.73 | 0.20 ± 0.19 | 0.79 ± 0.28 | -0.37 ± 0.14 | -0.55 ± 0.51 | 0.21 ± 0.12 |

**Supplementary Table 2** CO$_2$ fluxes (2015–2022 average, 2023 and H1-2024) for each ocean RECCAP2 region from bottom-up and inversion methods (2010–2022 from GCB 2024; 2023-2024 from NRT methods; Unit: GtC yr$^{-1}$).

| Period | Methods | World | Arctic | Atlantic | Coast | Indian | Pacific | Southern |
|---|---|---|---|---|---|---|---|---|
| 2015-2022 average | GOBMs | 2.44 ± 0.38 | 0.12 ± 0.03 | 0.51 ± 0.08 | 0.61 ± 0.11 | 0.27 ± 0.05 | 0.57 ± 0.12 | 1.06 ± 0.34 |
| | Data products | 2.52 ± 0.34 | 0.13 ± 0.04 | 0.56 ± 0.09 | 0.64 ± 0.12 | 0.18 ± 0.07 | 0.61 ± 0.13 | 0.99 ± 0.14 |
| | Inversions | 2.34 ± 0.45 | 0.09 ± 0.05 | 0.50 ± 0.13 | 0.48 ± 0.11 | 0.21 ± 0.09 | 0.50 ± 0.21 | 1.03 ± 0.17 |
| 2023 | GOBMs | 2.66 ± 0.50 | 0.10 ± 0.02 | 0.60 ± 0.13 | 0.66 ± 0.12 | 0.33 ± 0.11 | 0.73 ± 0.32 | 0.89 ± 0.40 |
| | Data products | 2.88 ± 0.28 | 0.11 ± 0.04 | 0.59 ± 0.08 | 0.64 ± 0.09 | 0.21 ± 0.05 | 0.69 ± 0.28 | 1.32 ± 0.09 |
| | Inversions | 2.26 ± 0.48 | 0.09 ± 0.02 | 0.46 ± 0.04 | 0.55 ± 0.09 | 0.18 ± 0.06 | 0.38 ± 0.15 | 1.14 ± 0.43 |
| H1-2024 | GOBMs | 2.82 ± 0.56 | 0.06 ± 0.01 | 0.88 ± 0.12 | 0.75 ± 0.14 | 0.14 ± 0.12 | 1.00 ± 0.37 | 0.54 ± 0.58 |
| | Data products | 3.32 ± 0.53 | 0.06 ± 0.03 | 0.73 ± 0.11 | 0.71 ± 0.13 | 0.14 ± 0.06 | 0.94 ± 0.38 | 1.58 ± 0.12 |
| | Inversions | 2.57 ± 0.31 | 0.06 ± 0.01 | 0.63 ± 0.14 | 0.64 ± 0.14 | 0.10 ± 0.08 | 0.44 ± 0.39 | 1.31 ± 0.46 |